\documentclass[journal=jctcce,manuscript=article]{achemso}
\usepackage[english]{babel}
\usepackage[utf8x]{inputenc}
\usepackage{amsmath}
\usepackage{graphicx}
\usepackage[T1]{fontenc}
\usepackage{array,calc,enumerate,booktabs}
\usepackage{multirow}
\usepackage{amssymb,amsmath,amstext,amsthm}
\usepackage[labelfont=bf,labelsep=period]{caption}
\usepackage[version=3]{mhchem}
\usepackage{graphicx,epstopdf}
\usepackage{setspace}
\usepackage{subcaption}
\usepackage{physics,color}

\newcommand{\comment}[1]{}

\newcommand*\citeref[1]{Ref~\citenum{#1}}
\newcommand*\citerefs[1]{Refs~\citenum{#1}}
\allowdisplaybreaks 

\title{CASSCF with Extremely Large Active Spaces using the Adaptive Sampling Configuration Interaction Method}
\author{Daniel S. Levine}
\affiliation[A]{Kenneth S. Pitzer Center for Theoretical Chemistry, Department of Chemistry, University of California, Berkeley, California 94720, USA}
\alsoaffiliation{Chemical Sciences Division, Lawrence Berkeley National Laboratory Berkeley, California 94720, USA}
\author{Diptarka Hait}
\affiliation[A]{Kenneth S. Pitzer Center for Theoretical Chemistry, Department of Chemistry, University of California, Berkeley, California 94720, USA} 
\alsoaffiliation{Chemical Sciences Division, Lawrence Berkeley National Laboratory Berkeley, California 94720, USA}
\author{Norm M. Tubman}
\affiliation[B]{Quantum Artificial Intelligence Lab (QuAIL), Exploration Technology Directorate, NASA Ames Research Center, Moffett Field, CA 94035, USA}
\author{Susi Lehtola}
\affiliation[C]{Chemical Sciences Division, Lawrence Berkeley National Laboratory Berkeley, California 94720, USA}
\alsoaffiliation{Present affiliation: Department of Chemistry, University of Helsinki, P.O. Box 55 (A. I. Virtasen aukio 1), FI-00014 University of Helsinki, Finland}
\author{K. Birgitta Whaley}
\affiliation[A]{Kenneth S. Pitzer Center for Theoretical Chemistry, Department of Chemistry, University of California, Berkeley, California 94720, USA}
\alsoaffiliation{Chemical Sciences Division, Lawrence Berkeley National Laboratory Berkeley, California 94720, USA}
\author{Martin Head-Gordon}
\affiliation[A]{Kenneth S. Pitzer Center for Theoretical Chemistry, Department of Chemistry, University of California, Berkeley, California 94720, USA} 
\alsoaffiliation{Chemical Sciences Division, Lawrence Berkeley National Laboratory Berkeley, California 94720, USA}
\email{mhg@cchem.berkeley.edu}
\begin{document}
\begin{abstract}
\textbf{Abstract}: The complete active space self-consistent field (CASSCF) method is the principal approach employed for studying strongly correlated systems. However, exact CASSCF can only be performed on small active spaces of {\raise.17ex\hbox{$\scriptstyle\sim$}}20 electrons in {\raise.17ex\hbox{$\scriptstyle\sim$}}20 orbitals due to exponential growth in the computational cost. We show that employing the Adaptive Sampling Configuration Interaction (ASCI) method as an approximate Full CI solver in the active space allows CASSCF-like calculations within chemical accuracy (<1 kcal/mol for relative energies) in active spaces with more than {\raise.17ex\hbox{$\scriptstyle\sim$}}50 active electrons in {\raise.17ex\hbox{$\scriptstyle\sim$}}50 active orbitals, significantly increasing the sizes of systems amenable to accurate multiconfigurational treatment. The main challenge with using any selected CI-based approximate CASSCF is the orbital optimization problem; they tend to exhibit large numbers of local minima in orbital space due to their lack of invariance to active-active rotations (in addition to the local minima that exist in exact CASSCF). We highlight methods that can avoid spurious local extrema as a practical solution to the orbital optimization problem. We employ ASCI-SCF to demonstrate lack of polyradical character in moderately sized periacenes with up to 52 correlated electrons and compare against heat-bath CI on an iron porphyrin system with more than 40 correlated electrons.
\end{abstract}

\section{Introduction}
Quantum chemical methods have advanced significantly for the treatment of most chemistry problems. Advances in density functional theory (DFT) have pushed the limits of system sizes to whole proteins\cite{fox2014density} and led to substantially improved accuracy across a spectrum of chemically relevant interactions\cite{mardirossian2017thirty,Goerigk2017GMTKN55,Najibi2018NLC,Mardirossian2018PT2,hait2018accurate,hait2018accuratepolar}. Local correlation methods\cite{Saebo1993local} for coupled-cluster theory have also dramatically increased the size of systems which can be treated accurately with wave function theory\cite{saitow2017new}. However, none of these methods are satisfactory for systems that display \textit{strong} correlation. Strong correlation is an effect that arises due to the presence of low-lying, accessible electronic states.\cite{Small2011SC} The quantum interference of these states with the single determinant ground state (implicitly or explicitly assumed in most flavors of DFT and coupled-cluster to be qualitatively correct) results in a wave function where many determinants have appreciable contribution.\cite{Shepard1987MC} Single-reference theories that ignore such effects introduce significant errors; indeed strong correlation is recognized as a primary failure mode of present-day DFT.\cite{mardirossian2017thirty} Strong correlation is commonly found in bond-breaking, extended $\pi$-systems, and (particularly first-row) transition metal systems\cite{hait2019levels}, especially those involving multiple metal centers. Dealing with strong correlation is thus extremely important for modeling catalysis, surface chemistry, and bioinorganic chemistry.

For decades, the standard method for addressing these sorts of problems has been the complete active space self-consistent field (CASSCF) method\cite{Roos1980CASSCF,siegbahn_comparison_1980,Siegbahn1981CASSCF,roos_complete_1987}. CASSCF is also sometimes known as the Fully Optimized Reaction Space (FORS) model.\cite{Ruedenberg1982FORS} In CASSCF, a subset of a system's orbitals and electrons are denoted \textit{active}, and the full configuration interaction (FCI) problem is solved exactly in this small active space. The remaining occupied orbitals are denoted \textit{inactive} and are treated in a mean-field manner, while the remaining unoccupied orbitals are denoted \textit{virtual}. The orbitals spanning these three spaces (inactive, active, and virtual) are then optimized to obtain the lowest possible energy. In other words, the CASSCF problem is to find the energy-optimal partitioning of the orbital Hilbert space. While the CASSCF reference captures strong correlation within the active space, it neglects weak or dynamic correlation, which is commonly described by either perturbation theory\cite{Andersson1992CASPT2,Roos1996CASPT2} or configuration interaction\cite{Harding2007reactive,Szalay2012MC} corrections.

CASSCF can be applied to moderately sized systems as long as the active space is relatively small, due to exponential growth in the number of possible Slater determinants that encompass all configurations within the active space. Indeed, the total number of possible Slater determinants for an active space with $M$ \textit{spatial} orbitals, $N_{\uparrow}$ up spins and $N_{\downarrow}$ down spins is
\begin{align}
 N_{Total}&=\binom{M}{N_{\uparrow}}\binom{M}{N_{\downarrow}}\label{combinatorial}
\end{align} 
Modern computing architectures can routinely handle active spaces of approximately 18 electrons in 18 orbitals ($\approx 2 \times 10^9$ determinants); massively parallel supercomputer implementations can partially converge results up to 22 electrons in 22 orbitals\cite{vogiatzis_pushing_2017} ($\approx 5\times 10^{11}$ determinants). However, these active space sizes are still not sufficient for studying many interesting problems in chemistry, especially those with multiple transition metal atoms. This has led to the development of physically motivated models to limit the number of determinants, such as the so-called restricted active space (RAS)\cite{Malmqvist1990RAS,Sherrill1999CI,Malmqvist2008RAS} and generalized active space (GAS)\cite{Ma2011GAS} variants of the CAS method. 

Another way to reduce the computational expense of CASSCF is to replace the exponentially scaling exact FCI solver with an approximate FCI method. Approximate FCI solvers that have been utilized to approximate CASSCF include Density Matrix Renormalization Group (DMRG)\cite{zgid2008density, ghosh2008orbital}%marti_density_2010}
, Full Configuration Interaction Quantum Monte Carlo (FCIQMC)~\cite{li_manni_combining_2016}, variational 2RDM~\cite{mazziotti_variational_2007,fosso2016large}, and incremental FCI~\cite{zimmerman_evaluation_2019}. In general, any approximate FCI solver can be employed to solve the FCI problem within an active space and thereby approximate CASSCF (as long as orbital gradients can be obtained).

A family of approximate FCI solvers with considerable promise in this regard are selected CI (SCI) methods. SCI methods are computationally advantageous under the assumption that only a \textit{relatively} small number of Slater determinants (out of the exponentially scaling total number given in Eqn \ref{combinatorial}) are actually important for even strongly correlated systems. These methods therefore attempt to identify (or "select") these important determinants via some criteria and diagonalize the wave function exactly within that reduced Hilbert subspace. An extra correction (such as from second order perturbation theory)\cite{huron1973} may potentially be added to approximate missing dynamic correlation. The general idea behind such methods is quite old\cite{bender1969,huron1973,evangelisti1983}, but the field has seen a considerable renaissance in recent years\cite{schriber2016communication,tubman2016-1,holmes2016,garniron2017} due to modern advances in computing that have made calculations with very large numbers of selected determinants feasible. SCI's ability to pick out important configurations makes it a natural candidate for black-box identification of determinants for general multiconfigurational self-consistent field (MCSCF) problems, with the results approaching the CASSCF limit with selection of increasing number of determinants. Indeed, an approximate CASSCF with the Heat-bath CI (HCI) solver has already been reported\cite{smith2017cheap}.

In this work, we present theoretical and practical details for employing the Adaptive Sampling Configuration Interaction (ASCI) method~\cite{tubman2016-1, tubman_modern_2018, tubman_efficient_2018} as the active space solver and carrying out a CASSCF procedure, which we term ASCI-SCF.
In the following sections, we briefly summarize these results and discuss particular details of the ASCI-SCF method arising from the unique nature of the ASCI wave function. We present data to show the benefits of orbital optimization for converging ASCI results, and demonstrate the ability of the ASCI-SCF to obtain CASSCF quality energies in systems of approximately 50 electrons in 50 orbitals.

\section{Theory}
\subsection{The ASCI Method}
\label{sec:asci_method}

The details of the ASCI method have already been presented elsewhere\cite{tubman2016-1, tubman_modern_2018, tubman_efficient_2018}, and so only provide a very brief summary will be provided. In the ASCI method, a trial CI wave function $\ket{\psi_k}$ is iteratively improved by the inclusion of new determinants (possibly replacing unimportant determinants) which are deemed important. The selection rule is derived from a consistency relationship among the CI coefficients of the exact FCI wave function. If we have an eigenstate wave function $\ket{\Psi}=\displaystyle\sum\limits_i C_i\ket{D_i}$ (where $\ket{D_i}$ are Slater determinants with coefficients $C_i$), then
\begin{equation}
C_{i} = \frac{\sum_{j \ne i}H_{ij}C_{j}}{H_{ii}-E}
\end{equation} 

where $H_{ij}=\bra{D_i}\mathbf{H}\ket{D_j}$ is the Hamiltonian matrix element between determinants $i$ and $j$, and $E$ is the energy of the eigenstate $\ket{\Psi}$. This exact relationship can be used as a metric to predict the expected weight of a determinant $\ket{D_i}$ in a CI expansion, by how it connects to other determinants in an approximate trial wave function. This connection is also used in Epstein--Nesbet perturbation theory~\cite{epstein1926,nesbet1955} for the coefficients of the determinants in the first order wave function.

In the ASCI method, all determinants $\ket{D_i}$ that are a single or double excitation from the most important determinants (as ranked by the magnitudes of their coefficients) in the trial wave function $\ket{\psi_k}$ are assigned an estimated importance $A_i$
\begin{equation}
A_{i} = \frac{\displaystyle \sum_{\ket{D_j} \in \ket{\psi_k}}H_{ij}C_{j}}{H_{ii}-E_k}
\end{equation}
where $E_k$ is the energy of the trial wave function $\ket{\psi_k}$. The search and selection is only done in the space spanned by determinants connected to the top $c$ determinants in $\ket{\psi_k}$ because unimportant determinants are unlikely to be the sole generators for a determinant with significant weight in the exact FCI wave function. This pruning of the search space greatly accelerates the algorithm. The top $t$ determinants (as ranked by magnitude of $A_i$) connected to $\ket{\psi_k}$ are then used to determine a new Hilbert subspace and, hence, a new wave function $\ket{\psi_{k+1}}$ by exact diagonalization within that subspace. 

Once several cycles of ASCI have been completed, the resulting wave function will contain all (or very nearly all) of the largest-weight determinants in the FCI wave function, while the determinants that have not been included should be of small weight in the exact FCI wave function. The effect of these many small remaining determinants are estimated by second-order Epstein--Nesbet perturbation theory~\cite{epstein1926,nesbet1955} (PT2). This final PT2 correction gives extremely accurate results, often within a kcal/mol of the absolute FCI energy even when only a tiny fraction of the Hilbert space is included in the ASCI wave function\cite{tubman2016-1,tubman_modern_2018,tubman_efficient_2018}. An extrapolation of the variational energy vs. the PT2 correction (to the FCI limit of of zero PT2 correction) can also be carried out to generate more accurate estimates, and predict a metric for error in the final estimate. It has been shown\cite{holmes2017excited,loos2018mountaineering,hait2019levels} that linear or quadratic fits are quite accurate for extrapolation of SCI energies against the PT2 correction. We observe essentially linear behavior of ASCI+PT2 energies\cite{hait2019levels}, and have consequently employed linear fits to refine results and estimate error (using the protocol described in \citeref{hait2019levels}).

\subsection{ASCI Orbital Gradients}
ASCI orbital gradients can be obtained via standard MCSCF procedures. Herein, we briefly recapitulate the MCSCF theory employed in ASCI-SCF orbital gradients.
\subsubsection{Density Matrices}
One- and two-particle density matrices (1-PDMs and 2PDMs, respectively) may be obtained from the ASCI wave function at a cost roughly comparable to a single Hamiltonian build. We have previously reported techniques for fast Hamiltonian construction via dynamic bitmasking \cite{tubman_modern_2018}, and the same techniques can also be employed to rapidly construct density matrices. In the following section, the permutational symmetry of 1- and 2-PDMs is elided and each symmetry related term is only given once. It is assumed in this work that both alpha and beta spin orbitals have the same spatial orbitals (that is, that the determinantal basis is restricted (possibly open-shell)). We follow the methods described by Helgaker, Jorgensen, and Olsen\cite{helgaker_configuration-interaction_2014}
. The one particle density matrix $D_{pq}$ can be written as the sum of two spin components $D^\alpha_{pq}$ and $D^\beta_{pq}$ which are given by
\begin{align}
    D_{pq}^\sigma = \bra{\Psi}E_{pq}^\sigma\ket{\Psi}
\end{align}
where $E_{pq}^\sigma$ is the spin-$\sigma$ excitation operator that excites from orbital $p$ to orbital $q$. If we denote a determinant by its alpha occupied orbitals $I_\alpha$ and beta occupied orbitals $I_\beta$ then this expression becomes 
\begin{align}
    D_{pq}^\sigma = \sum_{I_\alpha, I_\beta,J_\alpha, J_\beta}C_{I_\alpha I_\beta}\langle I_\alpha I_\beta|E_{pq}^\sigma|J_\alpha J_\beta\rangle C_{J_\alpha J_\beta}
\end{align}

It is useful to consider the density matrix in a determinant focused manner in a selected CI method, where the determinant list is not complete and $I_\alpha$ and $I_\beta$ are not independent, rather than in the typical orbital focused manner. Hence, we may re-express the above in terms of the contributions to $D_{pq}$ of pairs of determinants $I_\alpha I_\beta$ and $J_\alpha J_\beta$ that differ only by a single excitation from orbital $p$ to orbital $q$. This pair of determinants contributes $C_{I_\alpha I_\beta}C_{J_\alpha J_\beta}\gamma$ where $\gamma$ is a phase factor ($\pm 1$) due to the overlap of the bra and $p\rightarrow q$ excited ket. That is,
\begin{align}
\label{eq:opdm}
{}^{(I_\alpha I_\beta),(J_\alpha J_\beta)}D_{pq}&=C_{I_\alpha I_\beta}\langle I_\alpha I_\beta|E_{pq}^\sigma|J_\alpha J_\beta\rangle C_{J_\alpha J_\beta}\\
D_{pq} &= \sum_{\substack{(I_\alpha I_\beta),\\(J_\alpha J_\beta)}}{}^{(I_\alpha I_\beta),(J_\alpha J_\beta)}D_{pq}  
\end{align}

Similarly, the two-particle density matrix $\Gamma_{pqrs}$ % = \langle\Psi|E_{pq}-\delta_{rq}E_{ps}|\Psi\rangle$ 
can be written in terms of spin components
\begin{align}
\label{eq:tpdm}
\Gamma = \sum_{\substack{(I_\alpha I_\beta),\\(J_\alpha J_\beta)}}{}^{(I_\alpha I_\beta),(J_\alpha I_\beta)}\Gamma^\alpha + {}^{(I_\alpha I_\beta),(I_\alpha J_\beta)}\Gamma^\beta + {}^{(I_\alpha I_\beta),(J_\alpha I_\beta)}\Gamma^{\alpha\alpha} + {}^{(I_\alpha I_\beta),(I_\alpha J_\beta)}\Gamma^{\beta\beta} + {}^{(I_\alpha I_\beta),(J_\alpha J_\beta)}\Gamma^{\alpha\beta}    
\end{align}

The first term of \eqref{eq:tpdm} is non-zero for any pair of determinants that differ by exactly a single excitation (from $i$ to $a$) in the $\alpha$ space:
\begin{align}
\label{eq:tpdm-1}
    {}^{(I_\alpha I_\beta),(J_\alpha I_\beta)}\Gamma^\alpha_{pqrs} = (\delta_{pi}\delta_{qa}\delta_{rs}\delta_{r\in I_\alpha} - \delta_{pi}\delta_{qr}\delta_{r\in I_\alpha}\delta_{sa})C_{I_\alpha I_\beta}C_{J_\alpha I_\beta}\langle I_\alpha|E_{ia}^\alpha|J_\alpha \rangle
\end{align}
while the second term is its $\beta$ analogue. The third term is non-zero for any pair of determinants that differ by exactly a double excitation in the alpha space (from $i,j$ to $a,b$).
\begin{align}
\label{eq:tpdm-3}
   {}^{(I_\alpha I_\beta),(J_\alpha I_\beta)}\Gamma^{\alpha\alpha}_{pqrs} = (\delta_{pi}\delta_{qa}\delta_{rb}\delta_{sj} - \delta_{pi}\delta_{qb}\delta_{ra}\delta_{sj})C_{I_\alpha I_\beta}C_{J_\alpha I_\beta}\langle I_\alpha|E_{ia}^\alpha E_{jb}^\alpha|J_\alpha \rangle 
\end{align}

the fourth term again being the $\beta$ analogue. The final term is non-zero for any pair of determinants that differ by a single excitation in each of the $\alpha$ ($i$ and $a$) and $\beta$ ($j$ and $b$) spaces.
\begin{align}
    {}^{(I_\alpha I_\beta),(J_\alpha J_\beta)}\Gamma^{\alpha\beta}_{pqrs} = \delta_{pi}\delta_{qa}\delta_{rb}\delta_{sj}C_{I_\alpha I_\beta}C_{J_\alpha J_\beta}\langle I_\alpha|E^\alpha_{ia}|J_\alpha \rangle\langle I_\beta|E_{jb}^\beta|J_\beta \rangle
\end{align}

This gives the contributions of all singly- and doubly-connected determinant pairs to the 2-PDM in terms of CI coefficients and phase factors.

\subsubsection{Orbital Gradient}
Given the 1- and 2-PDMs, the generalized Fock matrices may be generated for any MCSCF. The derivation and further details are neatly described by Helgaker et.al.~\cite{helgaker_multiconfigurational_2014}, but the key results are summarized here. In the following, $m,n,p,q,\ldots$ are general indices, $i,j,k,\ldots$ are inactive indices, $t,u,v,w,\ldots$ are active indices, and $a,b,c,\ldots$ are virtual indices.

The generalized Fock matrix is defined as:
\begin{align}
  F_{mn}=\sum_{q}D_{pq}h_{pq} + \sum_{qrs}\Gamma_{mqrs}g_{nqrs}  
\end{align}
where $h_{pq}$ are the 1-electron integrals, $g_{nqrs}$ are the 2-electron integrals, and all indices run over all orbital classes (inactive, active, and virtual). This generally non-symmetric matrix can be simplified by taking advantage of the fact that the form of the density matrices are much simpler when some indices are inactive or virtual than when the indices are active. When the first index of the generalized Fock matrix is inactive and the second is general
\begin{align}
    F_{in} = 2({}^IF_{ni}+{}^AF_{ni})
\end{align}
where the \textit{inactive} and \textit{active Fock matrices} are, respectively, 
\begin{align}
{}^IF_{mn} &= h_{mn} + \sum_i(2g_{mnii}-g_{miin})\\
{}^AF_{mn} &= \sum_{vw}D_{vw}(g_{mnvw}-g_{mwvn})
\end{align}

In other words, the inactive Fock matrix is the Fock matrix formed from using only the inactive density and the active Fock matrix is sum of the Coulomb and exchange matrices built from the active space 1-PDM. When the first index is active, and the second index is general, we have
\begin{align}
F_{tn} = \sum_u {}^IF_{nu}D_{vu} + Q_{tn}    
\end{align}
where the auxiliary Q matrix is
\begin{align}
Q_{tm} = \sum_{u,v,w}\Gamma_{tuvw}g_{muvw}    
\end{align}
Finally, if the first index is virtual, then $F_{an} = 0$. This formulation of the generalized Fock matrix is useful because it only requires density matrices with all indices active and two-electron integrals in the MO basis with three indices active and one general index, greatly reducing the storage and computational cost of the MO transformation.

The orbital gradient is then given by
\begin{align}
    \frac{\partial E}{\partial \Delta_{pq}} = 2(F_{pq}-F_{qp})
\end{align}
One difference between the ASCI wave function and CAS wave functions is that active-active rotations are not generally redundant in the ASCI wave function. The question of the importance of these active-active rotations will be directly addressed in Section~\ref{aarots}.

\subsubsection{Hessian Preconditioner}
\label{sec:precond}
Optimization of the ASCI orbitals is carried out by a Broyden--Fletcher--Goldfarb--Shanno (BFGS) procedure, which yields quasi-Newton convergence behavior. In order to obtain rapid convergence, it is beneficial to precondition the gradient by the inverse of the Hessian matrix:
\begin{align}
    g' = H^{-1}g
\end{align}
Since inverting the Hessian matrix is computationally expensive, an approximate Hessian which is easily inverted (such as just the diagonal of the Hessian) is used in practice. Even constructing the exact Hessian diagonal can be too expensive, as the exact Hessian diagonal requires certain two electron integrals between orbitals in the spaces that are being rotated. Since we wish to carry out the MO transformation only in the active space, or with at most one general index, we neglect those integrals that have not been included in the MO transformation. In order to improve this approximation (neglecting these 2-electron integrals), we follow the suggestion of Chaban et al.~\cite{chaban_approximate_1997} and add small corrections to the inactive-active and active-virtual Hessian diagonal. The approximate Hessian diagonal elements are then
\begin{eqnarray}
&H_{ia,ia}& = 4({}^DF_{aa} - {}^DF_{ii})\\
&H_{ta,ta}& = 2D_{tt}{}^IF_{aa}-2\sum_u D_{tu}{}^IF_{tu} - 2\sum_{u,v,w} \Gamma_{tuvw}g_{muvw} + 2D_{tt}{}^AF_{aa}\\
&&= 2D_{tt}{}^IF_{aa}-2F_{tt} + 2D_{tt}{}^AF_{aa}\\
&&= 2D_{tt}{}^DF_{aa}-2F_{tt}\\
&H_{it,it}& = 4({}^DF_{tt} - {}^DF_{ii}) + 2D_{tt}{}^IF_{ii}-2\sum_u D_{tu}{}^IF_{tu} - 2\sum_{u,v,w} \Gamma_{tuvw}g_{muvw} + 2D_{tt}{}^AF_{ii}\\
&&= 2D_{tt}{}^IF_{ii}+4({}^DF_{tt} - {}^DF_{ii})-2F_{tt} + 2D_{tt}{}^AF_{ii}\\
&&= 2D_{tt}{}^DF_{ii}+4({}^DF_{tt} - {}^DF_{ii})-2F_{tt}
\end{eqnarray}

where ${}^DF = {}^IF + {}^AF$ and $F$ is the generalized Fock matrices defined above. In the case of the active-active orbital rotations, all of the necessary integrals have already been generated. Moreover, the active-active block is the most difficult block to converge due to the coupling of the active-active rotations to the CI expansion coefficients that are reoptimized after each orbital step by diagonalization of the Hamiltonian. Therefore, we use the exact active-active Hessian diagonal for preconditioning:
\begin{eqnarray}
H_{tu,tu} =&& \sum_{x,y}4 \Gamma_{txty}'(ux|uy)+4\Gamma_{uxuy}'(tx|ty)-8\Gamma_{txuy}'(tx|uy)\notag\\
&&\qquad +2 \Gamma_{ttxy}'(uu|xy)+2\Gamma_{uuxy}'(tt|xy)-4\Gamma_{tuxy}'(tu|xy)\notag\\
&&+2(D_{tt}{}^IF_{uu}+D_{uu}{}^IF_{tt}-2D_{tu}{}^IF_{tu}-F_{tt}-F_{uu})
\end{eqnarray}
where $\Gamma_{pqrs}' = \frac{1}{2}(\Gamma_{pqrs}+\Gamma_{pqsr})$.

\subsubsection{MCSCF Procedure}

The MCSCF employed in this work is of the \textit{inner-outer loop} variety, with the orbitals varied on the outer loop and the CI coefficients optimized on the inner loop. That is, the orbitals and CI coefficients are not optimized simultaneously; the orbitals are fixed, the CI coefficients are optimized with these orbitals, and then the orbitals are updated and fixed for a new CI optimization, and so on until convergence. Moreover, the CI determinant list as determined by ASCI is not altered during the MCSCF procedure. Rather, periodically (in the calculations described herein, every 20 MCSCF iterations), two cycles of ASCI updates are performed to ensure that the wave function still features the optimal determinants with the new orbitals. This strategy was chosen due to the strong coupling of the active-active rotations and the CI coefficients. The active-active rotations and CI coefficients contain redundancies that make their simultaneous optimization poorly conditioned~\cite{helgaker_multiconfigurational_2014}. This difficulty is avoided in CASSCF due to the fact that active-active rotations in this method are degenerate in energy and so are omitted from the MCSCF procedure. However, such rotations are non-degenerate in selected CI methods like ASCI-SCF and therefore must be included, necessitating an inner-outer loop structure to avoid this pitfall.

\subsection{ASCI Perturbation Theory}

After the ASCI method has achieved the desired convergence, the remainder of the Hilbert space (which by construction has low weight) is accounted for by second order perturbation theory (PT2). This PT2 correction is critical for obtaining FCI quality results as the cumulative effect of the relatively small, individual contributions from the many `not so important' determinants can be significant. However, the variationality of the energy has to be sacrificed for this improvement in accuracy. This therefore presents a question: for the purposes of optimizing orbitals, should the (variational) ASCI or ASCI+PT2 energy be optimized? We opt to only optimize the variational ASCI energy in this work. There are three principal reasons for this.

Including the PT correction in the orbital optimization should be most important when the underlying wave function being corrected is qualitatively wrong or of the wrong character. For example, in orbital-optimized M\o{}ller--Plesset perturbation theory (OO-MP2) vs non-orbital optimized variants (MP2), significant improvements in the final wave function are observed only when the Hartree--Fock reference on which the MP2 is based is qualitatively poor (for example, by being highly spin-contaminated); when the HF reference is good, orbital optimization does not substantially improve on non-optimized MP2\cite{neese_assessment_2009}. Since the purpose of the ASCI procedure is to obtain a qualitatively correct reference wave function, including the PT correction in the procedure is not expected to be significant; this expectation was borne out by recent work on approximate CASSCF with an HCI solver\cite{smith2017cheap}, where the presence of the PT2 terms was found to have minimal impact.

Secondly, by only optimizing a variational wave function, the optimization cannot be misled by any non-physical, non-variational behavior in the PT correction, which can become more pronounced if only a relatively small number of determinants are used in the ASCI wave function . Unregularized OOMP2 cannot dissociate bonds for instance, on account of the divergent nature of the perturbative corrrection\cite{stuck2013regularized,Lee2018kOOMP2} in the dissociation limit with restricted orbitals. Furthermore, orbital optimization that combines iterative amplitudes and perturbative amplitudes requires careful regularization of the perturbative amplitudes.\cite{Lawler2008penalty} 

Finally, orbital gradients are significantly less complicated and much less computationally expensive when only the variational wave function is optimized. Since the PT corrections are not expected to be significant, we believe this high cost cannot be justified. A final PT correction to the orbital optimized variational ASCI wave function should however still be calculated in the end and extrapolations performed when necessary.

\section{Results}

We now demonstrate some of the power of the ASCI-SCF method as applied to a diverse set of organic and inorganic molecules.

\subsection{The importance of active-active orbital rotations}\label{aarots}
We begin with a somewhat technical question: are the active-active rotations in ASCI-SCF important to include as optimization parameters? In principle, if the ASCI wave function was converged exactly (that is, as the ASCI wave function approaches the FCI wave function), active-active orbital rotations would not change the energy and therefore be completely ignorable. However, this level of convergence is never achieved in practice for systems where the exact FCI problem is intractable, and so the active-active rotations must be considered. It is well known that truncated CI expansions predict lower variational energies when using approximate natural orbitals than when canonical Hartree--Fock orbitals are employed \cite{szabo_modern_1996}, and similar behavior has been observed for SCI wave functions. Specifically, we have previously shown that rotating from canonical HF orbitals to the natural orbitals obtained by forming and diagonalizing the active space 1-PDM substantially improves the convergence of the ASCI wave function by creating a more compact representation in which lower variational energy may be obtained with the same number of determinants~\cite{tubman2016-1,tubman2018postponing}, but can this be further improved by using the 2-PDM and two-electron integrals to find orbital gradients?

\begin{table}[htb!]
\begin{tabular}{cccccc}
\cline{1-6}
&\multirow{2}{*}{ASCI (100k)}&\multirow{2}{*}{ASCI-SCF (100k)}&\multirow{2}{*}{ASCI (300k)}&SCF stab.&Incr. det. stab.\\
&&&&(kcal/mol)&(kcal/mol)\\
\cline{1-6}
\multirow{2}{*}{cc-pVDZ} & & & & & \\
 & & & & & \\\cline{1-6}
H$_3$COH&-115.40054&-115.41209&-115.40857&7.24&5.03\\
CH$_3$Cl&-499.42603&-499.43680&-499.43185&6.75&3.65\\
H$_3$CSH&-438.03796&-438.04838&-438.04498&6.54&4.41\\
HOCl&-535.22099&-535.23033&-535.22647&5.86&3.43\\
SO$_2$&-547.68359&-547.69205&-547.69587&5.31&7.70\\
%N2H4&-111.54561&-111.55333&-111.55379&4.85&5.13\\
%H2O2&-151.17999&-151.18557&-151.18564&3.50&3.55\\
%ClO&-534.57320&-534.57797&-534.57778&2.99&2.87\\
%ClF&-559.19411&-559.19829&-559.19771&2.62&2.26\\
%H2CO&-114.21265&-114.21653&-114.21694&2.43&2.69\\
%C2H6&-79.55671&-79.55990&-79.56623&2.00&5.97\\
%SO&-472.66310&-472.66616&-472.66763&1.92&2.85\\
%S2&-795.33304&-795.33590&-795.33839&1.79&3.35\\
%Si2H6&-581.59617&-581.59893&-581.60259&1.73&4.03\\
%CS&-435.60590&-435.60861&-435.60934&1.70&2.16\\
%P2&-681.73137&-681.73384&-681.73467&1.55&2.07\\
\cline{1-6}
\multirow{2}{*}{cc-pVTZ} & & & & & \\
 & & & & & \\\cline{1-6}
 H$_3$CSH&-438.13456&-438.18038&-438.16427&28.75&18.65\\
CO$_2$&-188.26289&-188.30784&-188.29965&28.20&23.06\\
SO$_2$&-547.88864&-547.92974&-547.92515&25.79&22.91\\
CH$_3$Cl&-499.54311&-499.58109&-499.56638&23.83&14.60\\
N$_2$H$_4$&-111.64221&-111.67997&-111.67282&23.69&19.21\\
Si$_2$H$_6$&-581.71766&-581.75427&-581.74293&22.97&15.86\\
C$_2$H$_6$&-79.62516&-79.66140&-79.65307&22.74&17.51\\
H$_3$COH&-115.51465&-115.54928&-115.53570&21.73&13.21\\
Cl$_2$&-919.42110&-919.44838&-919.44523&17.12&15.14\\
HOCl&-535.38388&-535.41065&-535.40107&16.80&10.79\\
ClO&-534.73118&-534.75322&-534.75092&13.83&12.39\\
P$_2$&-681.83866&-681.85976&-681.85962&13.24&13.16\\
ClF&-559.38152&-559.40175&-559.39619&12.70&9.21\\
H$_2$O$_2$&-151.32851&-151.34654&-151.34623&11.31&11.12\\
H$_2$CO&-114.32288&-114.34089&-114.33844&11.30&9.76\\
\cline{1-6}
\end{tabular}
\caption{ASCI and ASCI-SCF variational energies with 100000 determinants and ASCI variational energies with 300000 determinants for the G1 test set (in a.u.) in the cc-pVDZ basis, sorted by the magnitude of the SCF stabilization (i.e. difference between the ASCI and ASCI-SCF energies). The non ASCI-SCF calculations employ approximate natural orbitals for wave function compaction. The SCF stabilization and stabilization obtained by tripling the number of determinants (in kcal/mol) are given in the last two columns. Only cases with SCF stabilization > 5 kcal/mol have been reported herein (full listing for the entire G1 set available in Supporting Information).}
\label{table:DZ}
\end{table}

In order to assess the improvement of the ASCI variational wave function, orbital optimization was carried out on the G1 test set alongside approximate FCI calculations with ASCI, within the cc-pVDZ and cc-pVTZ basis sets. As can be seen in the partial data in Table~\ref{table:DZ} (full results available in the Supporting Information), active-active orbital optimization lowers the energy of the variational wave function by a similar amount as tripling the size of the wave function (and sometimes significantly more), particularly when the wave function is relatively far from convergence at the ASCI level. In particular, up to 7 kcal/mol of stabilization in the case of cc-pVDZ and up to almost 30 kcal/mol in the case of cc-pVTZ can be obtained by active-active rotations. This substantial variational improvement creates a much more compact wave function, thereby improving the accuracy with the same sized wave function expansion and reducing the cost to investigate large systems of interest. Specifically, lower variational energies lead to a reduction in the magnitude of the non-variational PT2 correction and thereby enhances confidence in the quality of the results. This becomes especially evident if extrapolation is carried out, as smaller PT2 corrections are closer to the $E_\text{PT2}=0$ limit, leading to smaller scope for substantial deviation (and thereby predicting smaller error, as described in \citeref{hait2019levels}). 

\begin{table}[htb!]
\begin{tabular}{lll}
\cline{1-3}
Method             & \multicolumn{2}{l}{Energies (in Ha)} \\
                   & Approx. Natural orbitals        & ASCI-SCF           \\
\cline{1-3}
ASCI (variational) & -1545.32054       & -1545.34729      \\
PT2                & -0.08503       & -0.05597      \\
ASCI+PT2           & -1545.40558       & -1545.40326      \\
Extrapolated       & -1545.444(11)      & -1545.4184(8)  \\   
\cline{1-3}
\end{tabular}
\caption{Energy of NiC/def2-TZVPP with 2 million variational determinants and various orbital representations. The approximate natural orbitals were computed with 2 million variational determinants, while the ASCI-SCF orbitals were optimized with 50000 variational determinants. The extrapolation employs a least-squares fit to the 500000, 1, and 2 million determinant results.}
\label{tab:nic}
\end{table}

As a practical demonstration, we present an attempt to get the CASCI energy of the difficult transition metal diatomic carbide NiC, within the def2-TZVPP\cite{weigend2005balanced} basis set, employing a six-orbital frozen core (corresponding to 1s of C and 1s, 2s, and 2p of Ni), resulting in a (22e, 89o) active space. Table~\ref{tab:nic} details the energies obtained with approximate natural orbitals and ASCI-SCF optimized orbitals (with active-active rotations only). It can be seen that a substantial reduction in variational energy (by roughly 16 kcal/mol) is obtained with ASCI-SCF orbitals (utilizing only 50000 determinants) relative to approximate natural orbitals obtained from 2 million determinants. The larger PT2 correction for the non-orbital optimized calculation brings the two ASCI+PT2 energies to much closer agreement (of approx 1 kcal/mol). However, the sheer magnitude of the large PT2 correction erodes the reliability of the orbital unoptimized result, as an extrapolation to the CASCI limit of zero PT2 correction (in the manner described in \citeref{hait2019levels}) indicates very large error bars. On the other hand, chemical accuracy can be attained from the ASCI-SCF orbitals, as they have rather small PT2 corrections (on account of having more negative variational energies), leading to much lower extrapolation error.
In general therefore, ASCI-SCF orbital rotations could prove very useful in reducing the error in the extrapolated energy, by pushing more correlation energy into the fully reliable variational component.

\subsection{How polyradical are periacenes?}

\begin{figure}[htb!]
\includegraphics[width=0.5\textwidth]{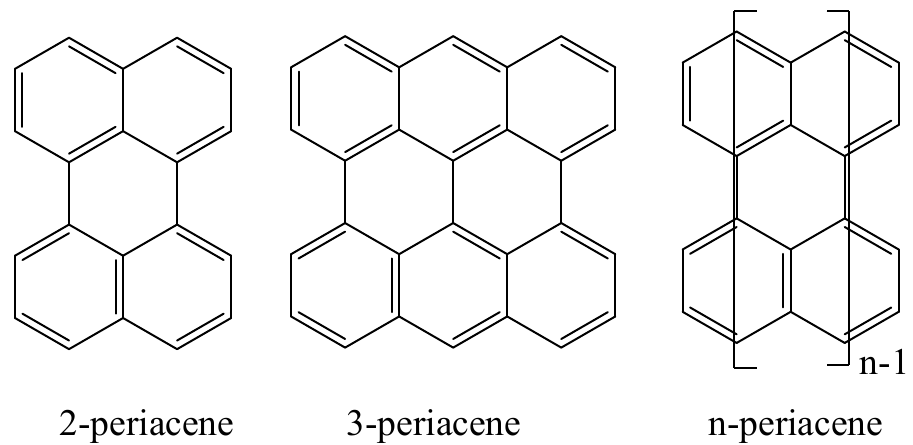}
\caption{Molecular structure of periacenes. 2-periacene is more commonly known as perylene.}
\label{fig:periacenes}
\end{figure}

Periacenes are two-dimensional analogues of acenes, where two parallel acene chains are directly bonded together to create a 2D sheet (as shown in Fig \ref{fig:periacenes}). The electronic structure of periacenes is potentially interesting as they could serve as cluster models for graphene. In particular, delocalization of $\pi$ electrons over the 2D sheet could result in a polyradicaloid singlet ground state with large numbers of effectively unpaired electrons (as has been suggested to be the case for 1D acenes\cite{hachmann2007,mizukami2013}). A number of theoretical studies\cite{mizukami2013,rivero2013entanglement,mullinax2019heterogeneous,plasser2013multiradical} have consequently been performed on periacenes to estimate the onset of ground state polyradical character, making them ideal systems for determining the utility of the ASCI-SCF method.

\begin{table}[]
\begin{tabular}{llllllll}
\cline{1-8}Periacene & Active Space & \multicolumn{2}{c}{Wall times} & \multicolumn{2}{c}{Singlet Energy}                                  & \multicolumn{2}{c}{Singlet-Triplet gaps}                  \\ 
& & \multicolumn{2}{c}{(in s)} & \multicolumn{2}{c}{ (in Ha.)}                                  & \multicolumn{2}{c}{ (in kcal/mol)}                  \\\cline{1-8}
          &              & CI                    & OO                    & This work      &  {v2RDM}  & This work & {v2RDM} \\
2         & (20e,20o)    & 100                   & 42                    & -764.70604(9)  & -764.78776                                                & 38.3(1)   & 35.4                                                        \\
3         & (28e,28o)    & 80                    & 80                    & -1068.9736(3)  & -1069.1184                                                & 17.5(2)   & 19.5                                                        \\
4         & (36e,36o)    & 80                    & 119                   & -1373.2514(8)  & -1373.4549                                                & 10.0(6)   & 13.4                                                        \\
5         & (44e, 44o)    & 110                   & 212                   & -1677.5198(15) & -1677.7927                                                & 4 (1)%3.6 (13) 
& 10.8                                                        \\
6         & (52e,52o)    & 140                   & 351                   & -1981.7642(9)  & -1982.1298                                                &  8 (1)%8.2 (12) 
& 8.9                                                        
\end{tabular}
\caption{Absolute ground state singlet energies and adiabatic singlet-triplet gaps for periacenes, as estimated by extrapolated ASCI+PT2 using ASCI-SCF orbitals.  Comparisons have been made to the v2RDM results from \citeref{mullinax2019heterogeneous}. Wall timings (for both CI and OO (orbital optimization) steps) per ASCI-SCF iteration are also provided. Timings were obtained on 24 cores of an AMD EPYC 7401 (2.0 GHz, 64MB cache) machine for calculations using 1 million variational determinants.}
\label{tab:periaceneenergies}
\end{table}

We have carried out ASCI-SCF orbital optimizations on 2-6 periacenes on the full valence $\pi$ subspace (i.e. the $2p_z$ orbital and one electron per C atom), using v2RDM-CASSCF optimized geometries from \citeref{mullinax2019heterogeneous}, the cc-pVDZ basis set and 1 million variational determinants for the wave function. Extrapolated ASCI+PT2 energies using the converged ASCI-SCF orbitals were subsequently computed (as reported in Table \ref{tab:periaceneenergies}) to estimate the true CASSCF energies (with the largest ASCI+PT2 calculation utilized in the extrapolation for each species employing $\ge $ 5 million determinants). This extrapolation is required in order to obtain high quality results. Two electron integral construction for the orbital optimization process was accelerated with the RI approximation\cite{feyereisen1993use,weigend1998ri} (using the rimp2-cc-pVDZ auxiliary basis set\cite{weigend1998ri}), but the reported final ASCI+PT2 results (using ASCI-SCF optimized orbitals) do not employ this approximation. Wall times for individual ASCI-SCF cycles have also been reported in Table \ref{tab:periaceneenergies}, which highlights the ease of running such calculations on medium sized computers. 

Table \ref{tab:periaceneenergies} compares our results to those of \citeref{mullinax2019heterogeneous} (which employs the same geometries and basis set). The level of qualitative agreement for the singlet-triplet gaps is quite decent, with the exception of the increase in the gap moving from 5-periacene to 6-periacene with our method.
There is, however, a fair bit of quantitative disagreement between the two approaches, well beyond the error
bars predicted by extrapolated ASCI+PT2. Part of this might originate from the systematic underestimation of absolute energies by the v2RDM approach employed in \citeref{mullinax2019heterogeneous} due to underconstraint of the 2RDM. Indeed, Table \ref{tab:periaceneenergies} shows that the absolute energies of singlet states of 2-periacene differ by $\approx 0.08$ a.u. between v2RDM-CASSCF and extrapolated ASCI+PT2, and larger differences are seen for larger systems (roughly increasing by $0.06$ a.u. per each step on the periacene sequence). The small active space size of 2-periacene however seems to suggest that a selected CI approach is likely to be rather effective in estimating the true CASSCF energy for this problem, and the small estimated extrapolated ASCI+PT2 error bar of $\approx 10^{-4}$ a.u. appears to support this viewpoint. On the other hand, the v2RDM values result from calculations employing 2 body PQG N-representability constraints\cite{garrod1964reduction} alone and are not systematically improvable without imposition of more stringent N-representability constraints. This is quite unlike ASCI, where systematic improvement can always be obtained via increasing the size of the variational subspace. We are therefore more inclined to trust the ASCI values at present, but comparison with v2RDM employing tighter N-representability constraints (such as the three body constraints) could prove interesting. Most of the systematic difference in absolute energy is cancelled in relative energies like singlet-triplet gaps at any rate, leading to much smaller differences in those values between our work and \citeref{mullinax2019heterogeneous}. It is similarly worth noting that most of the remaining size-consistency CI errors (that were not captured by PT2) would likely systematically cancel for energy gap computations. Such errors however would be strictly smaller than the error against the true CASCI values, and the estimated errors in Table \ref{tab:periaceneenergies} suggest that size-consistency errors are not a significant issue for the reported gaps. 

The one anomalous case is the increase in singlet-triplet gap on moving from 5-periacene to 6-periacene with our method, which appears to be counterintuitive (relative to the expectation that the gap would monotonically decay with system size). There exists a possibility that our calculations reached a local extrema in orbital space (although every effort was made to avoid such extrema, as detailed in Sec \ref{extrema}), leading to a spurious ordering of singlet-triplet gaps. However, approximate (Yamaguchi) spin-projected\cite{yamaguchi_spin_1988} DFT calculations, which should be quite accurate for biradical species, as these periacenes are found to be (\textit{vide infra}) predict the same behavior (see Supporting Information). This suggests that the anomalous behavior is more likely either reflective of exact quantum mechanics or is an artifact associated with the mismatch between the levels of theory used to compute the geometries and the singlet-triplet gaps. At any rate, similar behavior has been observed for linear acenes in the past\cite{fosso2016large,ghosh2017generalized,dupuy2018fate}. Further investigations using different geometries and orbital guesses for these species might be necessary in order to determine if this behavior is indeed real or an artifact induced either by the geometry employed or by our method. The relatively small size of the energy gaps involved however makes this quite a challenging task to definitively settle.

\begin{table}[htb!]
\begin{tabular}{llllll}
\cline{1-6}
       & 2-periacene & 3-periacene & 4-periacene & 5-periacene & 6-periacene \\
       \cline{1-6}
LUNO+1 & 0.10      & 0.10      & 0.11  & 0.11  & 0.10      \\
LUNO   & 0.19      & 0.25      & 0.56   & 0.87  & 0.92      \\
HONO   & 1.82      & 1.75      & 1.44  & 1.13  & 1.08      \\
HONO+1 & 1.90      & 1.90      & 1.89  & 1.89  & 1.91      \\
\cline{1-6}
\end{tabular}
\caption{Frontier natural orbital occupations of the singlet state for the periacene sequence, from ASCI wave functions with 5 million determinants. Here HONO and LUNO stand for Highest Occupied and Lowest Unoccupied Natural Orbitals respectively (i.e. the N$_e$ and N$_e$+1 natural orbitals, ordered by occupancy), following earlier literature \cite{mizukami2013,rivero2013entanglement}.}
\label{tab:natocc}
\end{table}

The rather small singlet-triplet gap for the larger periacenes appears to indicate emergence of radicaloid character in the singlet. Indeed, \citerefs{rivero2013entanglement,mullinax2019heterogeneous,plasser2013multiradical} suggest that varying degrees of polyradicaloid character emerge by 6-periacene. We however only see substantial evidence of biradical character, and very little evidence for polyradical character along the periacene sequence up to 6-periacene. Frontier natural orbital (NO) occupations of these species with an ASCI solution using 5 million determinants are reported in Table \ref{tab:natocc}, which shows that 5- and 6-periacenes are essentially biradicaloid, with HONO and LUNO occupations close to unity. The LUNO+1 and HONO-1 however do not appear to have occupations characteristic of strong unpairing (i.e. do not substantially deviate from 0 or 2 along the periacene sequence), suggesting lack of polyradical character. It must be noted that the ASCI wave function itself does not converge as quickly as the energy, as changes in the variational wave function only appear as second-order perturbations of  the energy, and perturbation expansions and extrapolations have not been carried out for the wave function. However, significant deviations from the closed-shell NO occupations of $0$ or $2$ should correspond to important degrees of freedom that ASCI is quite good at selecting\cite{tubman_modern_2018,tubman_efficient_2018} out of the full Hilbert space. The absence of dynamical correlation contributions from the ASCI wave function therefore ought not to strongly affect the NO occupations corresponding to these degrees of freedom. Indeed, the frontier NO occupations were relatively quite stable for various ASCI variational wave function subspace sizes, suggesting overall lack of polyradical character even if the NO occupations themselves are not as fully converged as the energy. Interestingly, the 6-periacene has a slightly larger singlet-triplet gap than 5-periacene, despite being slightly more biradical (based on NO values in Table \ref{tab:natocc}).

\begin{table}[htb!]
\begin{tabular}{llllll}
\cline{1-6}
Acene  & Top  & Top 2 & Top 3 & Top 4 & Top 5 \\
\cline{1-6}
2 & 0.66 & 0.67  & 0.68  & 0.68  & 0.69  \\
3 & 0.58 & 0.62  & 0.63  & 0.63  & 0.64  \\
4 & 0.45 & 0.60  & 0.61  & 0.62  & 0.62  \\
5 & 0.33 & 0.57  & 0.58  & 0.59  & 0.60  \\
6 & 0.30 & 0.55  & 0.55  & 0.56  & 0.57  \\
\cline{1-6}
\end{tabular}
\caption{Fraction of the singlet ASCI wave function cumulatively contributed by the $N$ top contributing determinants ($N=1-5$), using an ASCI wave function with 5 million determinants.}
\label{tab:overlaps}
\end{table}

Furthermore, we can attempt to determine what fraction of the ASCI-SCF wave function stem from the contributions of only a few determinants\cite{tubman2018postponing}, in an attempt to understand the structure of the full CASSCF wave function. This should be reasonably accurate, as the ASCI selection rule is rather effective at selecting the most important contributors to the full wave function\cite{tubman_modern_2018,tubman_efficient_2018}. Table \ref{tab:overlaps} shows the cumulative contributions of the top $N$ determinants ($N=1-5$) for the singlet periacenes. We can see that while the contribution of the top determinant (the HF-like determinant with aufbau filling) decreases with increasing system size, going from 66\% in the 2-periacene to 30\% in the 6-periacene, the top two determinants collectively contribute approximately $55-65\%$ for \textit{all periacenes}. The third, fourth and fifth top determinants collectively contribute only $2-3\%$ to the total wave function for all periacenes as well, showing that the importance of these determinants does not grow with system size. While a quantitative distinction between contributors to static and dynamic correlation cannot yet be cleanly drawn based on the determinant contributions, it is quite clear that the top two determinants are the only ones with considerable ($>10\%$) weight in the full wave function. The multireference character of the periacenes therefore appears to principally be a two determinant problem, which suggests undeniable biradical character for the higher periacenes but essentially no polyradical nature. Yamaguchi approximate spin-projected DFT\cite{yamaguchi_spin_1988} and related methods therefore have the potential to be reasonably accurate for such systems, supporting our earlier observations regarding the anomalous singlet-triplet gap increase on going from 5-periacene to 6-periacene.

The difference between our observations and predictions of polyradical character from previous studies can perhaps be understood by noting that both the SUHF method employed in \citeref{rivero2013entanglement} and the v2RDM method used in \citeref{mullinax2019heterogeneous} tend to predict ``more radical-like" NO occupations than CASSCF. At any rate, prior experience with 1D polyacenes suggest that radicaloid character is artificially augmented by pure $\pi$ space calculations, and inclusion of $\sigma$ orbitals into the active space reduces radical character\cite{lee2017coupled,lehtola2018orbital,schriber2018combined}. It is therefore reasonable to view these $\pi$ space calculations as an upper bound to the true radicaloid character of these molecules, and our results suggest that members of the periacene sequence up to 6-periacene are at best biradical and possess very little polyradical character. 

\subsection{ The case of Iron Porphyrin}
\begin{figure}[htb!]
\includegraphics[width=0.5\textwidth]{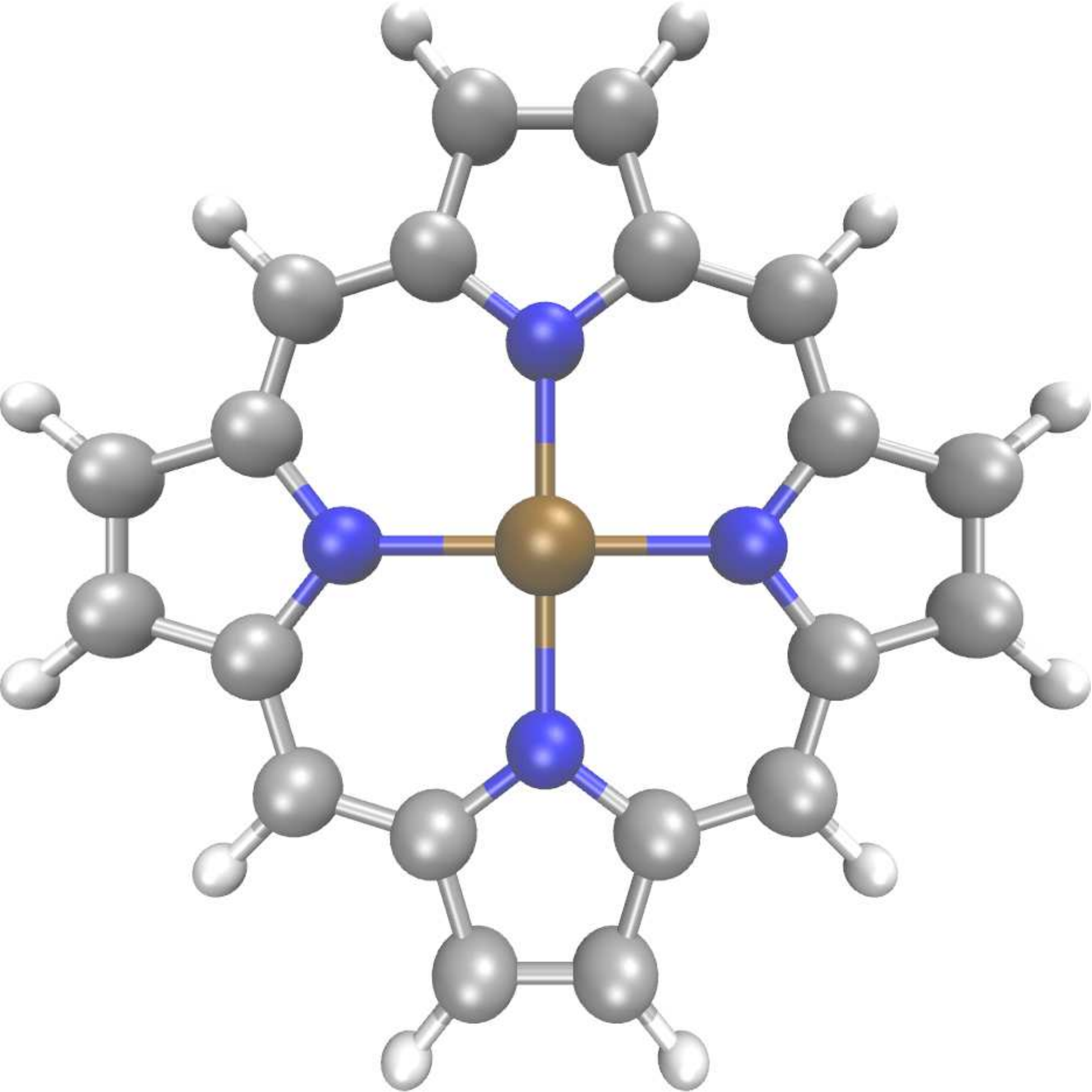}
\caption{Model system for Iron porphyrin. Gray atoms are C, white is H, blue is N and brown is Fe.}
\label{fig:porphyrin}
\end{figure}
\subsubsection{Comparison between ASCI-SCF and HCISCF}
Previous studies employing large active space methods, including DMRG\cite{olivares-amaya_ab-initio_2015}, FCI-QMC\cite{li_manni_combining_2016,li2018understanding,li2019role}, and HCISCF\cite{smith2017cheap} have been applied to understand the electronic structure and theoretical ground spin state of iron porphyrin. Experimental investigations of Fe(II) porphyrin\cite{collman1975synthesis,mispelter1980proton, lang1978mossbauer} have found a triplet ground state, but theoretical studies (on the model system depicted in Fig \ref{fig:porphyrin}) have often found a quintet ground state\cite{li_manni_combining_2016,choe1998theoretical,choe1999theoretical,radon2008binding,phung2016cumulant}, though some studies do report a triplet ground state when an extremely large active space is used\cite{olivares-amaya_ab-initio_2015,smith2017cheap,lee2020utilizing}. \citeref{smith2017cheap} in particular presents a detailed HCISCF study of this system, making it a natural point for comparing and contrasting the ASCI-SCF method against HCISCF.

Like \citeref{smith2017cheap}, we have performed our calculations on the optimized triplet geometry presented in \citeref{groenhof2005electronic}, employing the cc-pVDZ basis and HF orbitals of the quintet state as the initial guess. \citeref{smith2017cheap} examines two active spaces for this species: a ``small'' one with 32 electrons in 29 orbitals and a larger one with 44 electrons in 44 orbitals. The first corresponds to a selection of the five Fe $3d$ orbitals and the 24 $\pi$ orbitals of the porphyrin ring, while the larger active space also includes the five Fe $4d$ orbitals (to account for any double-shell effect), the Fe $4p_x$ and $4p_y$ orbitals and the eight N $2p_x$ and $2p_y$ orbitals. It must however be noted that the orbital optimization procedure can pick out an even more optimal set of orbitals by pushing out weakly correlated orbitals out of the active space in favor of more relevant ones.

\begin{table}[]
\begin{tabular}{lllll}
\hline
Active space &     State    & Method           & N$_\text{DETS}$ & E$_\text{CI}$   \\ \hline
(32e, 29o)    & $^5A_{1g}$ & ASCI-SCF (small) & 100000     & -2245.0096 \\
             &         & vHCISCF  ($^5A_g$)         & 379536     & -2244.9980 \\
             &         & ASCI-SCF  & 500000     & -2245.0191 \\
             & $^3A_{2g}$ & ASCI-SCF (small) & 100000     & -2244.9913 \\
             &         & vHCISCF ($^3B_{1g}$)          & 533623     & -2244.9776 \\
             &         & ASCI-SCF  & 500000     & -2245.0022 \\
             &         &                  &            &            \\
(44e, 44o)    & $^5A_{1g}$ & ASCI-SCF (small) & 100000     & -2245.2073 \\
             &         & vHCISCF ($^5A_g$)          & 1450271    & -2245.1457 \\
             &         & ASCI-SCF  & 500000     & -2245.2315 \\
             & $^3A_{2g}$ & ASCI-SCF (small) & 100000     & -2245.1822 \\
             &         & vHCISCF  ($^3B_{1g}$)         & 2133424    & -2245.1567 \\
             &         & ASCI-SCF  & 500000     & -2245.2044 \\ \hline
\end{tabular}
\caption{Comparison of variational energies E$_\text{CI}$ from ASCI-SCF and variational HCISCF (vHCISCF) for Fe-porphyrin/cc-pVDZ, respectively. N$_\text{DETS}$ is the total size of the variational subspace. ASCI-SCF state symmetries were based on $d$ orbital occupations, and are somewhat approximate on account of symmetry breaking in the (44e, 44o) active space. HCISCF energies and state symmetries have been taken from \citeref{smith2017cheap} (which appears to assume a $D_{2h}$ point group for symmetry assignments despite using a $D_{4h}$ geometry). The difference between the variational energies suggests that the two approaches have converged onto different orbitals.}
\label{tab:porph}
\end{table}

Table \ref{tab:porph} compares the variational, orbital optimized energies obtained from ASCI-SCF and HCISCF. It appears that ASCI-SCF is able to obtain substantially lower variational energies with smaller variational subspaces (by a factor of 4-20) than HCISCF for this system. The HCISCF results therefore appear to correspond to spurious local extrema in orbital space, as opposed to the true CASSCF global minimum. While we cannot prove that the ASCI-SCF solutions are the true global minima either, they are considerably ``better" in a variational sense and have significantly lower energy even after inclusion of PT2 corrections and extrapolation (as can be seen from Table \ref{tab:pt2energies}). As an extreme example, we can see that the ASCI-SCF variational energies for the (44e, 44o) active space (given in Table \ref{tab:porph}) go below the extrapolated SHCI (HCI with stochastic PT2) energies (given in Table \ref{tab:pt2energies}), let alone the purely variational HCI estimate! An immediate implication is that ASCI-SCF predicts a quintet ground state for both active spaces, in direct contrast to HCISCF (which predicted a triplet ground state for the larger active space). Therefore, the conclusions in \citeref{smith2017cheap} regarding porphyrin should not be viewed as definitive as they appear to stem from HCISCF finding a local extremum as a solution.

There are two (major) possible explanations for why HCISCF yields higher energy solutions than ASCI-SCF: HCI may fail to select important configurations (over less important ones) due to its use of a more approximate selection rule or the orbital optimization algorithm may be less efficient (\citeref{smith2017cheap} does not mention any preconditioners, for one). These are the main differences between the ASCI-SCF and HCISCF approaches; ASCI yields more compact wave functions (i.e. lower energy for same number of determinants) than HCI\cite{tubman_efficient_2018,tubman_modern_2018} and the ASCI-SCF orbital optimization uses a very reasonable preconditioner (described in Sec \ref{sec:precond}), which likely assists in avoiding some local extrema in orbital space. It is difficult to determine the extent to which either factor is individually responsible for the discrepancy, especially since the rather problematic nature of this active space leads to a proliferation of local extrema (as discussed below).

\begin{table}[]
\begin{tabular}{llll}
\hline
Active space &        & Extrapolated ASCI+PT2 & Extrapolated SHCI \\
\hline
(32e, 29o)    & $^5A_{1g}$ & -2245.03241(3)        & −2245.0314(5)     \\
             & $^3A_{2g}$ & -2245.01865(1)       & −2245.0049(6)     \\
             &         &                       &                   \\
(44e, 44o)    & $^5A_{1g}$ & -2245.28688(7)         & −2245.1964(9)     \\
             & $^3A_{2g}$ & -2245.25617(8)         & −2245.1995(6) \\
             \hline
\end{tabular}
\caption{Comparison of extrapolated E$_\text{CI}$+E$_\text{PT2}$ from ASCI (using ASCI-SCF orbitals) and SHCI (using vHCISCF orbitals) for Fe-porphyrin/cc-pVDZ, respectively. SHCI values have been taken from \citeref{smith2017cheap}.}
\label{tab:pt2energies}
\end{table}

\begin{figure}[htb!]
\begin{minipage}{0.48\textwidth}
\includegraphics[width=\linewidth]{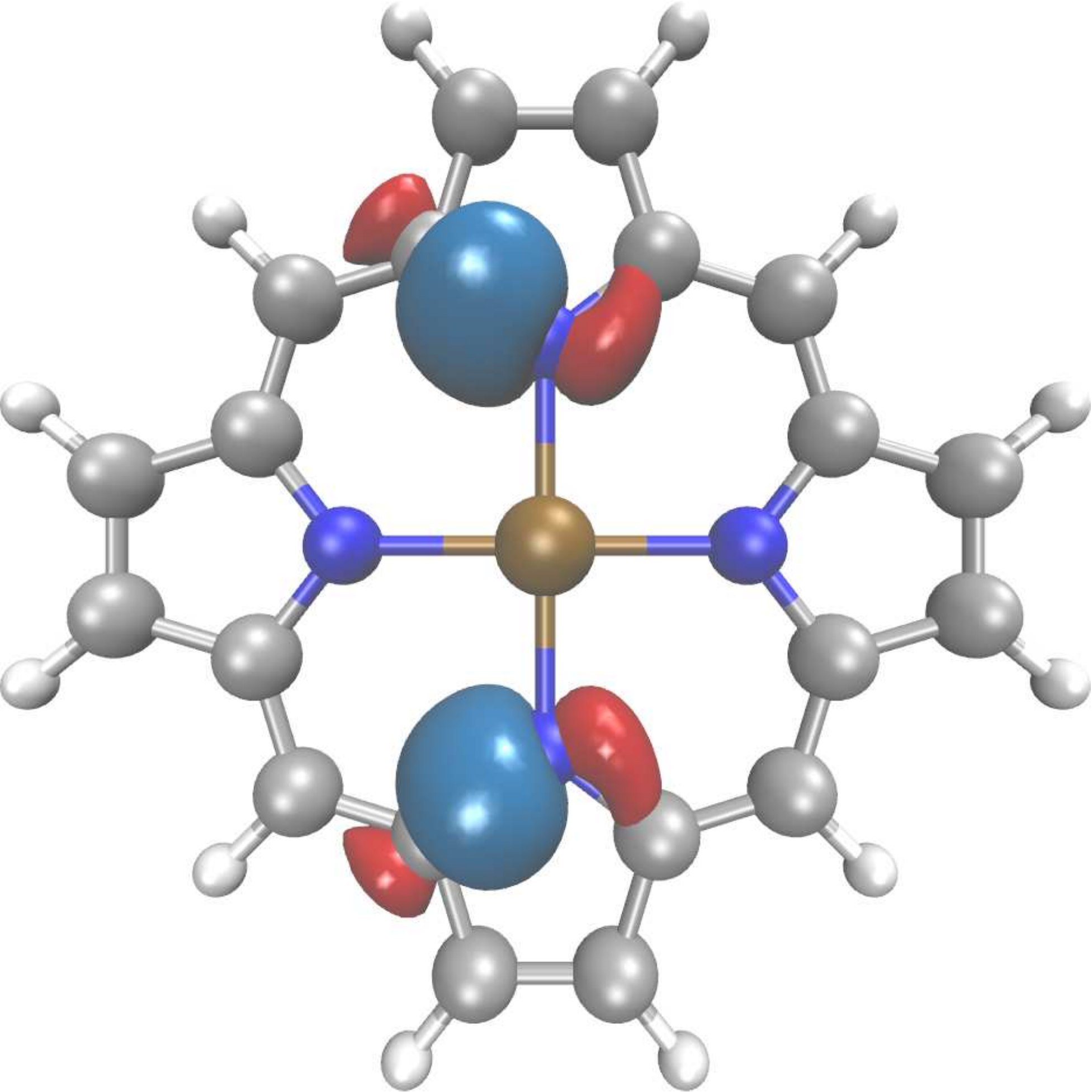}
\subcaption{$\sigma$}
\end{minipage}
\begin{minipage}{0.48\textwidth}
\includegraphics[width=\linewidth]{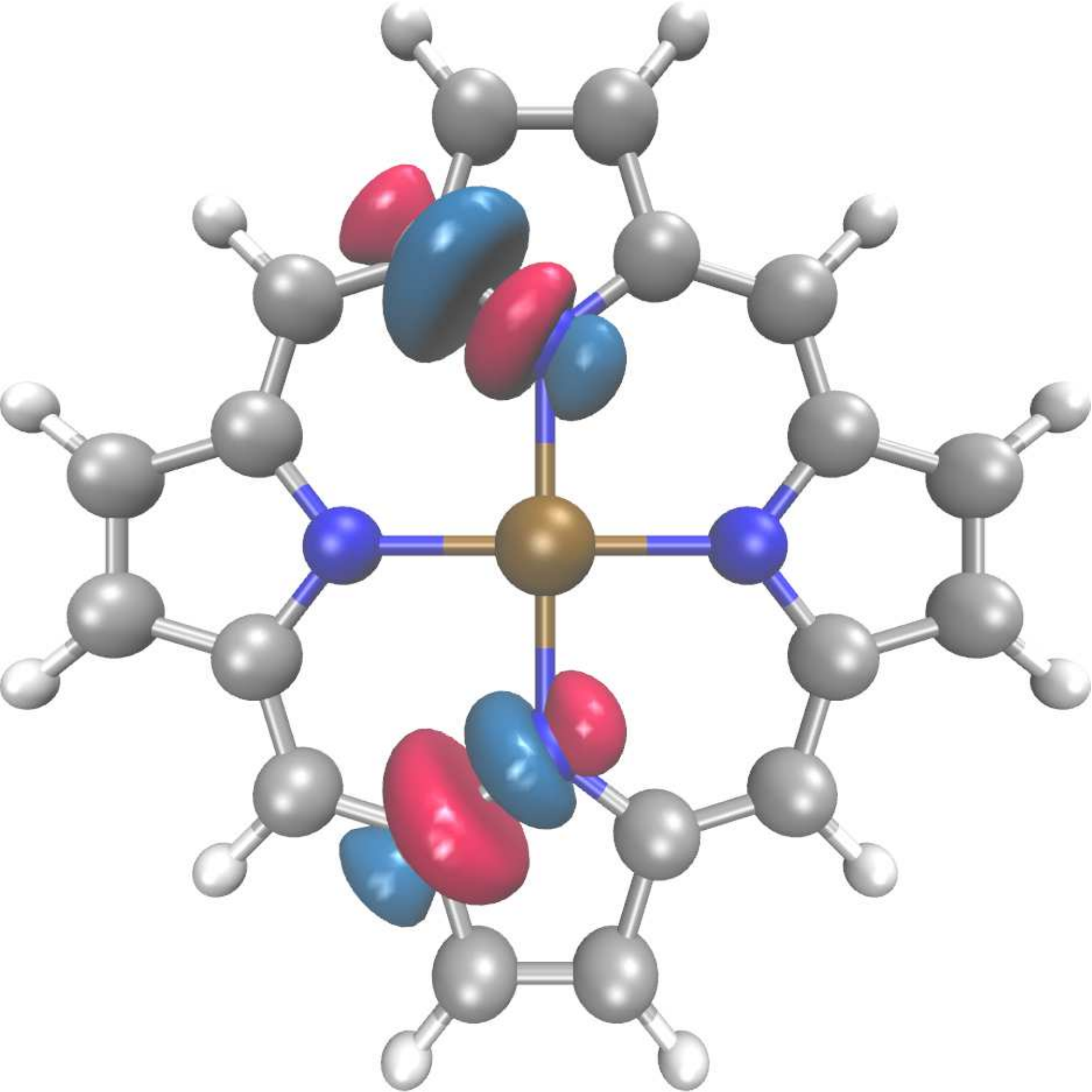}
\subcaption{$\sigma^*$}
\end{minipage}
\caption{Some ASCI-SCF C-N $\sigma$ orbitals for the $^3A_{2g}$ state optimized with the (44e, 44o) active space. Note that spatial symmetry has been broken in favor of localization.}
\label{fig:unbalanced}
\end{figure}

The large quintet-triplet gap of $\approx 19$ kcal/mol predicted by ASCI-SCF with the (44e, 44o) active space was nonetheless quite troubling as the states appeared to be fairly single-reference (based on the natural orbital occupations) and many single reference methods predict much smaller gaps (irrespective of the spin of the ground state)\cite{groenhof2005electronic,lee2020utilizing}. We consequently examined the ASCI-SCF orbitals, and discovered that this active space of 44 electrons in 44 orbitals was likely unbalanced. The proposed active space composition includes eight 2$p_{\{x,y\}}$ orbitals of N that are involved in $\sigma$ bonding. Four of these orbitals are involved in coordination to the metal and are consequently likely to be important. However, the other four are involved with C-N $\sigma$ bonding, in conjunction with the $2s$ orbital. In other words, it is problematic to include only the $p$ orbitals in the $sp^2$ hybridized N centers, since the exclusion of the $s$ orbitals would lead to an unbalanced number of C-N $\sigma/\sigma^*$ orbitals entering the active space. The exclusion of the C $sp^2$ orbitals involved in this bonding is also problematic. Overall, the (44e, 44o) active space includes two $\sigma$ and two $\sigma^*$ C-N orbitals out of the 16 possible ones. The remaining C-N $\sigma/\sigma^*$ orbitals are proximate in energy to the four already in the active space, and are likely to correlate together on account of their locality. Inclusion of all of these orbitals would subsequently raise the question of whether C-C $\sigma/\sigma^*$ orbitals should be considered or not; such lines of questioning can quickly get out of hand. 

Indeed, ASCI-SCF pushes out relatively less correlated orbitals like the $4d_{x^2-y^2}$ in order to incorporate more C-N $\sigma/\sigma^*$ orbitals into the active space instead. Examples of some resulting ASCI-SCF orbitals can be seen in Fig \ref{fig:unbalanced}. We note that this behavior occurs \textit{despite} ensuring that the initial active space has only orbitals of the right symmetry, indicating that it is energetically more favorable to incorporate extra C-N $\sigma/\sigma^*$ orbitals in the active space than include some weakly interacting ones that are tangentially involved with the metal ligand interaction. Exchanging the final orbitals between triplet and quintet states (in order to have a consistent set) does not lead to any further lowering of energy (in fact, the energies either increase considerably or return to previous values), nor does exchanging orbitals in/out of the active space prior to re-optimization lead to any further stabilization. This active space therefore appears to be fundamentally problematic and prone to local extrema based on the number and nature of C-N $\sigma/\sigma^*$ orbitals that creep into the active space.  Checkpoint files for all these orbitals have been provided in the Supporting Information for further examination by interested readers.

\subsubsection{Behavior of a 40 electron, 42 orbital active space}

\begin{figure}[htb!]
\begin{minipage}{0.24\textwidth}
\includegraphics[width=\linewidth]{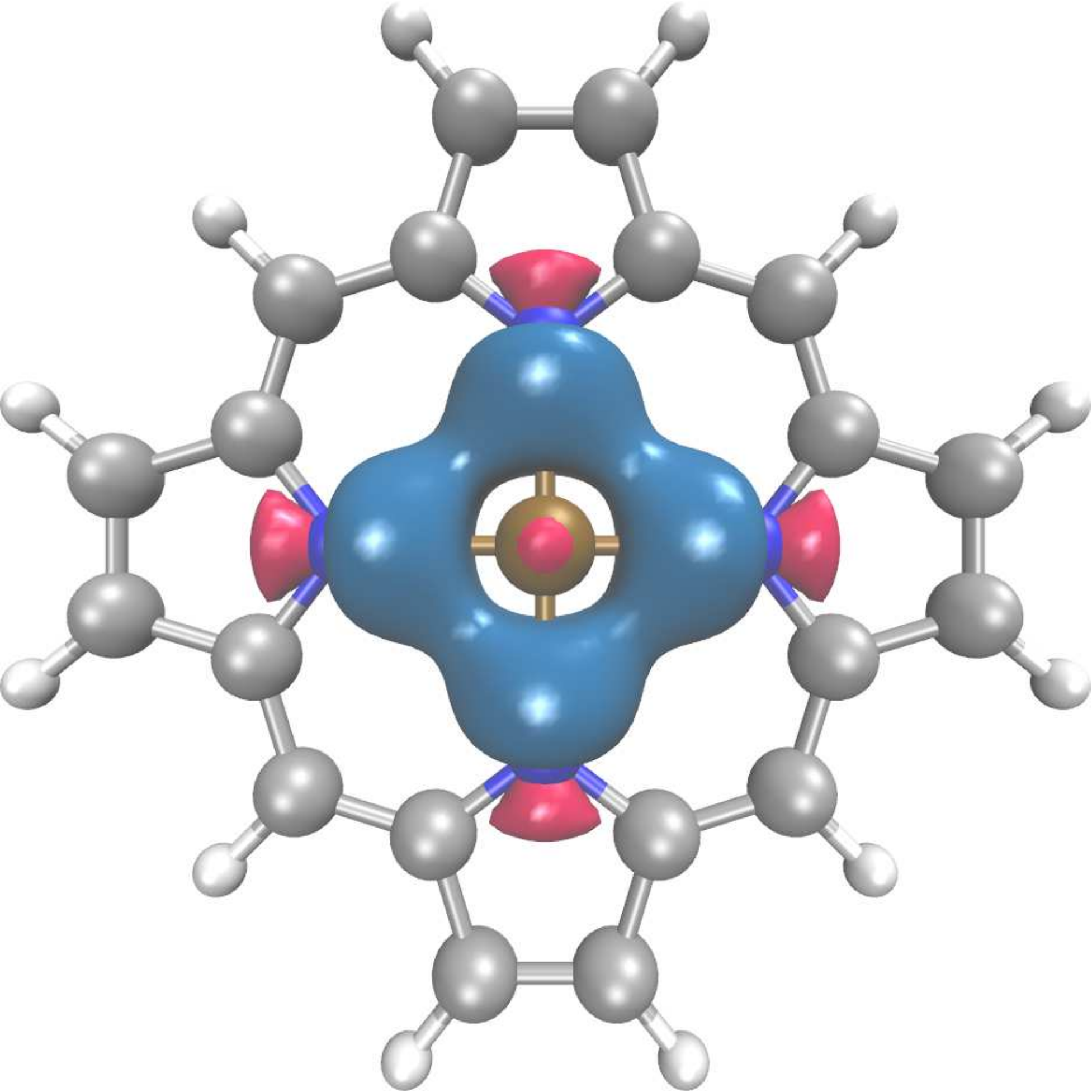}
\end{minipage}
\begin{minipage}{0.24\textwidth}
\includegraphics[width=\linewidth]{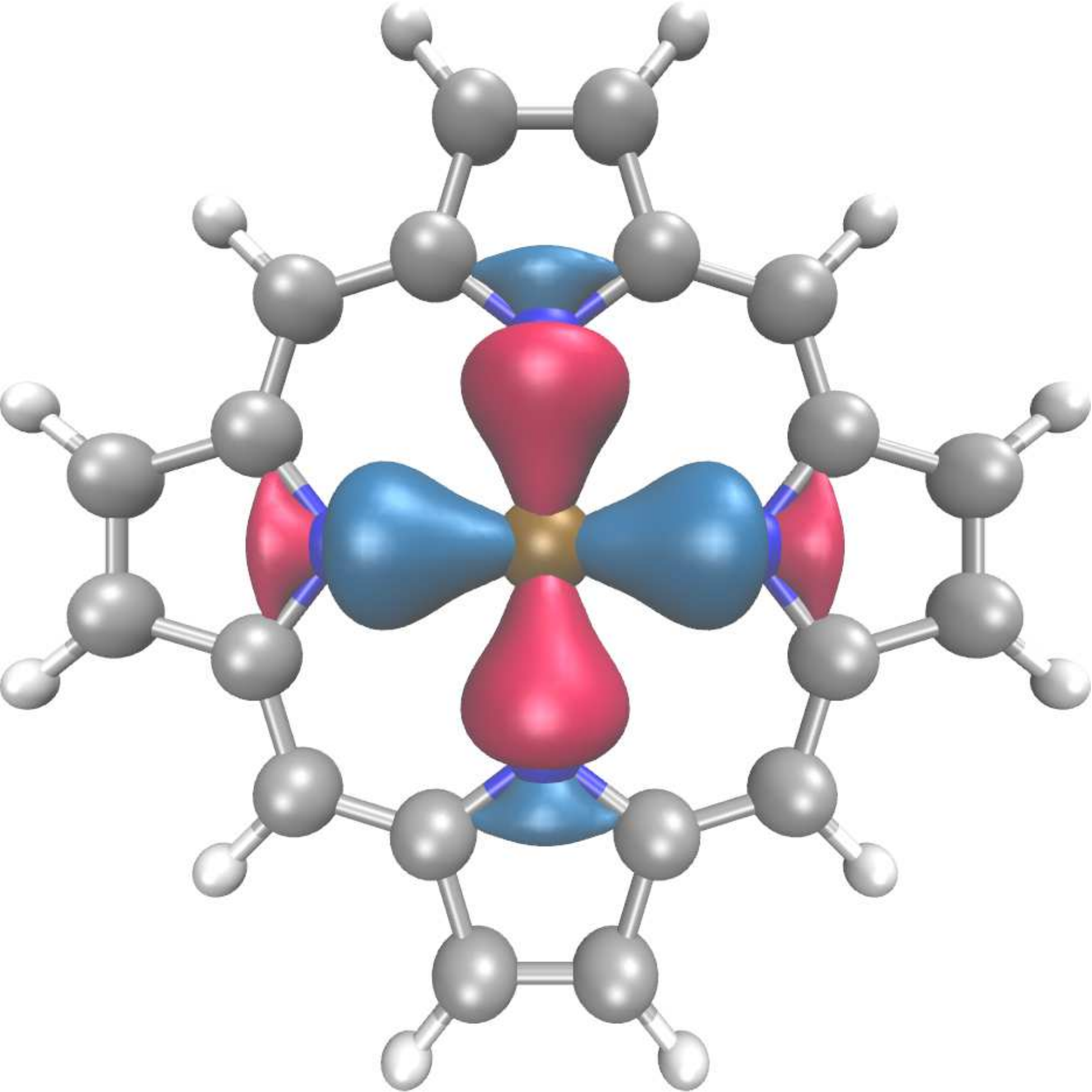}
\end{minipage}
\begin{minipage}{0.24\textwidth}
\includegraphics[width=\linewidth]{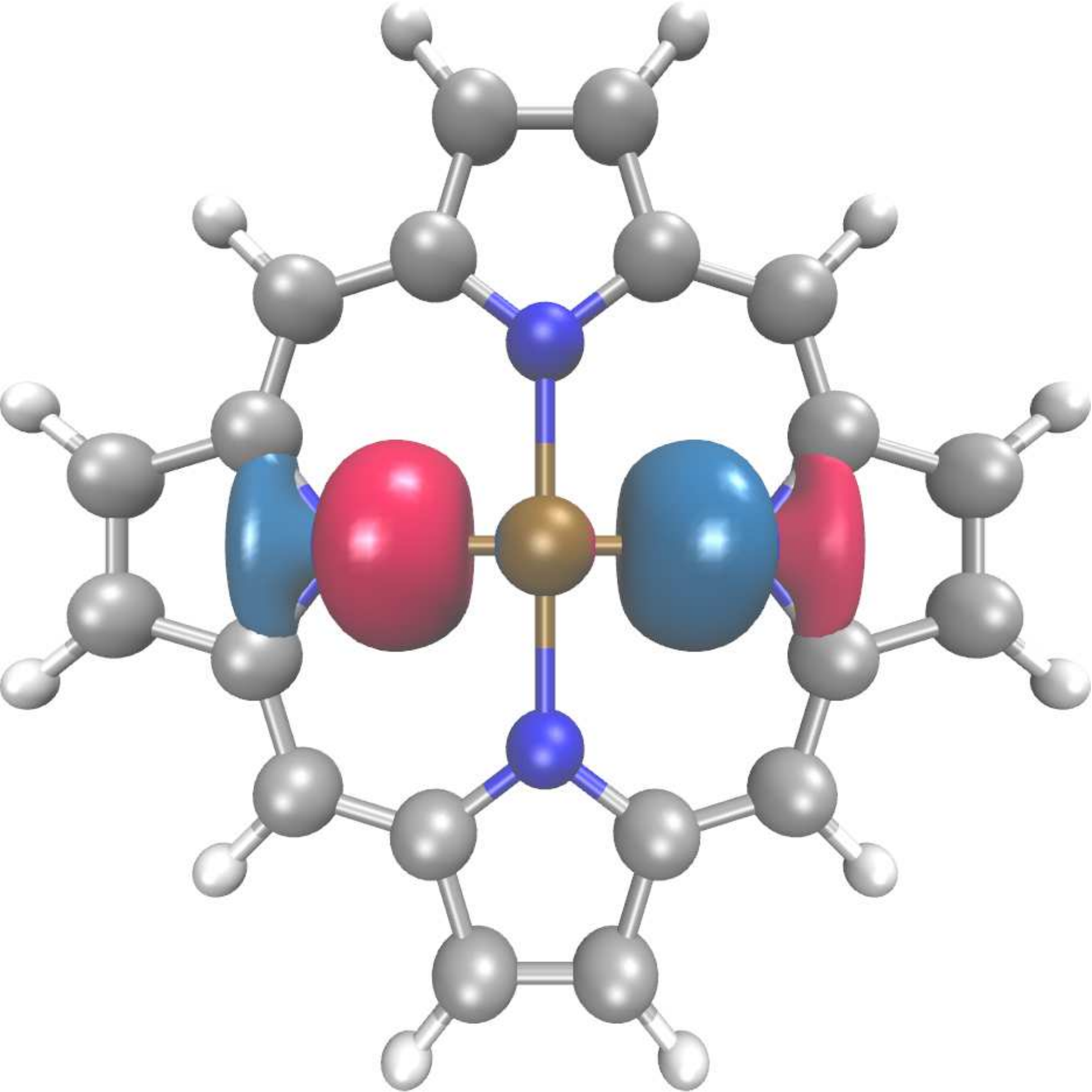}
\end{minipage}
\begin{minipage}{0.24\textwidth}
\includegraphics[width=\linewidth]{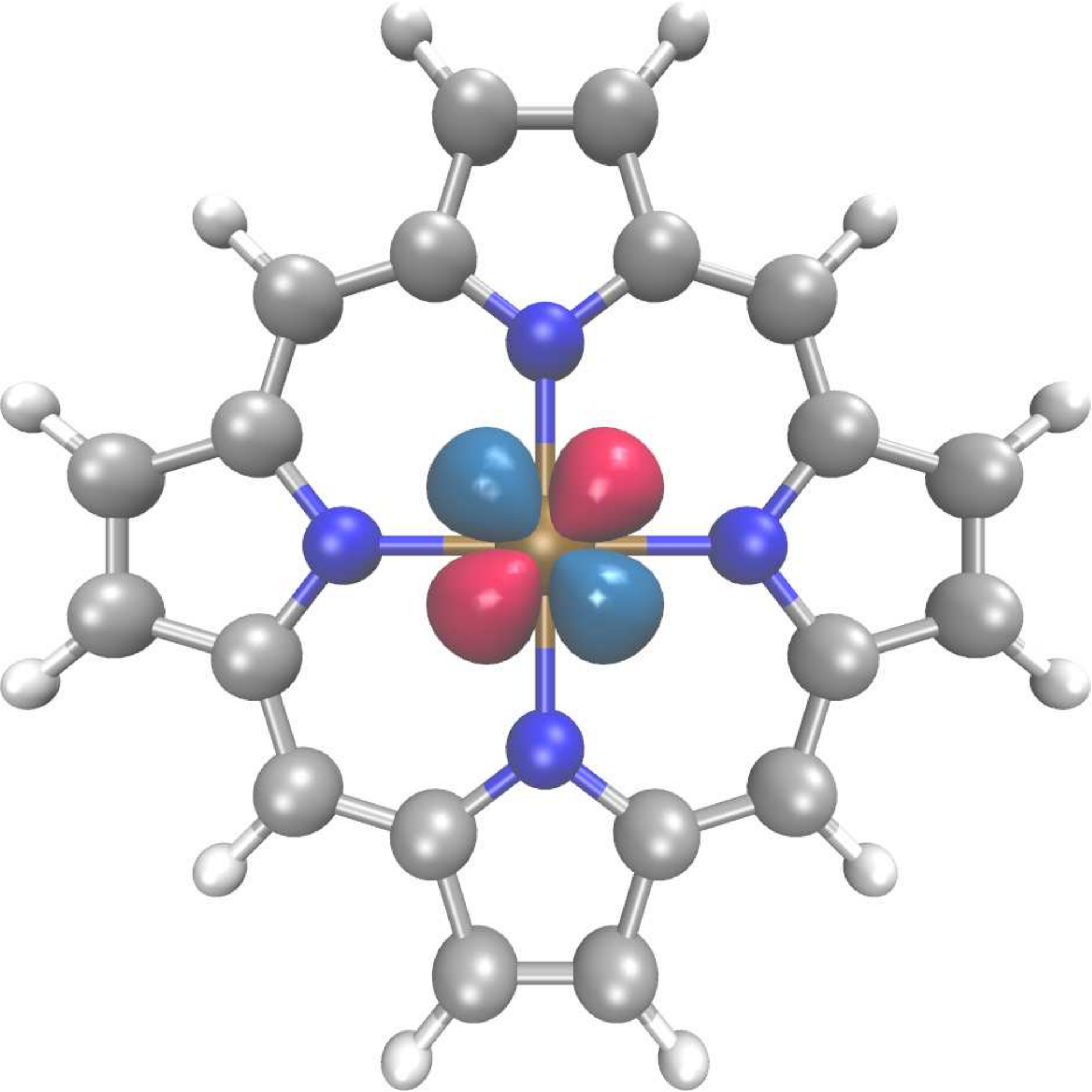}
\end{minipage}
\begin{minipage}{0.24\textwidth}
\includegraphics[width=\linewidth]{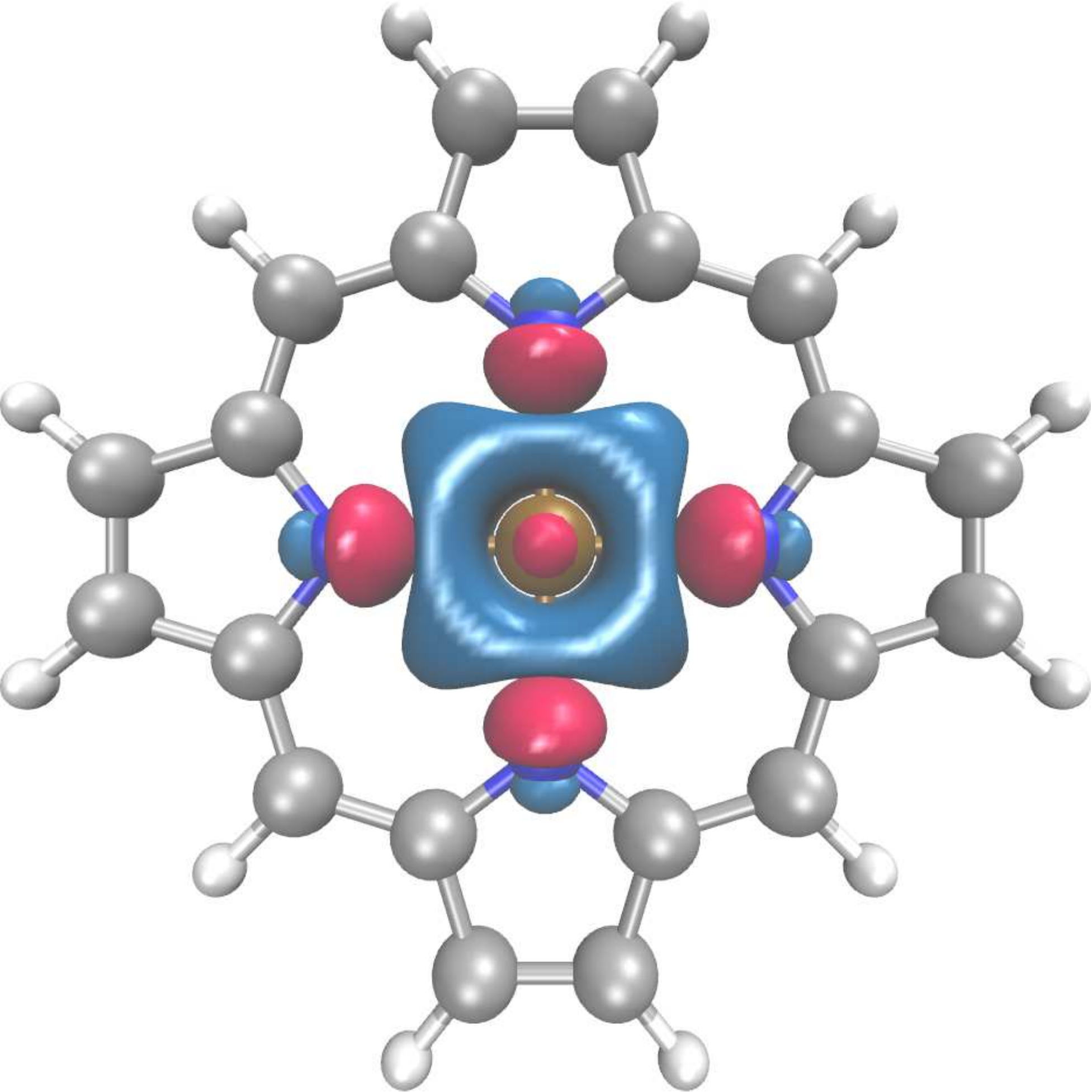}
\subcaption{$a_{1g}$}
\end{minipage}
\begin{minipage}{0.24\textwidth}
\includegraphics[width=\linewidth]{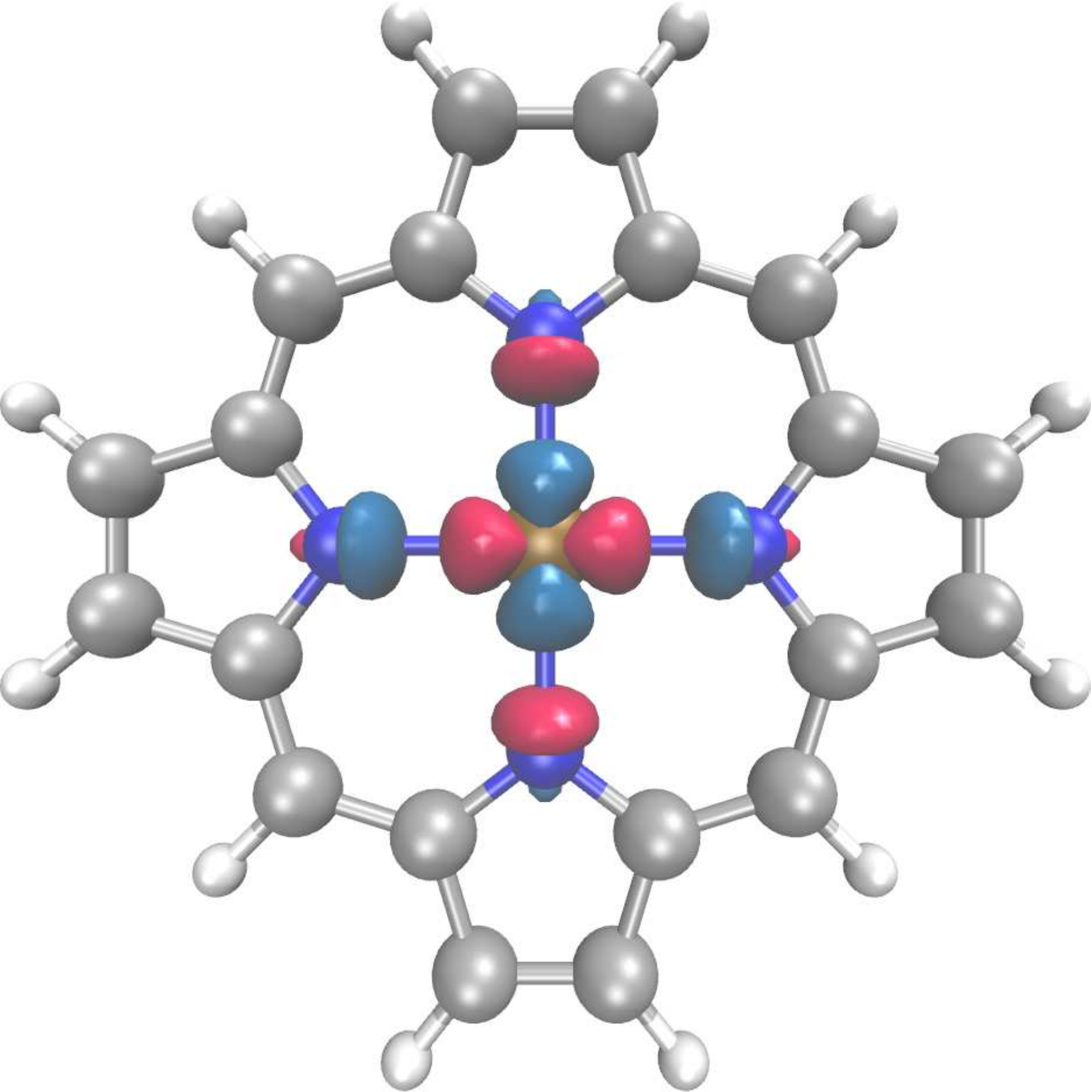}
\subcaption{$b_{1g}$}
\end{minipage}
\begin{minipage}{0.24\textwidth}
\includegraphics[width=\linewidth]{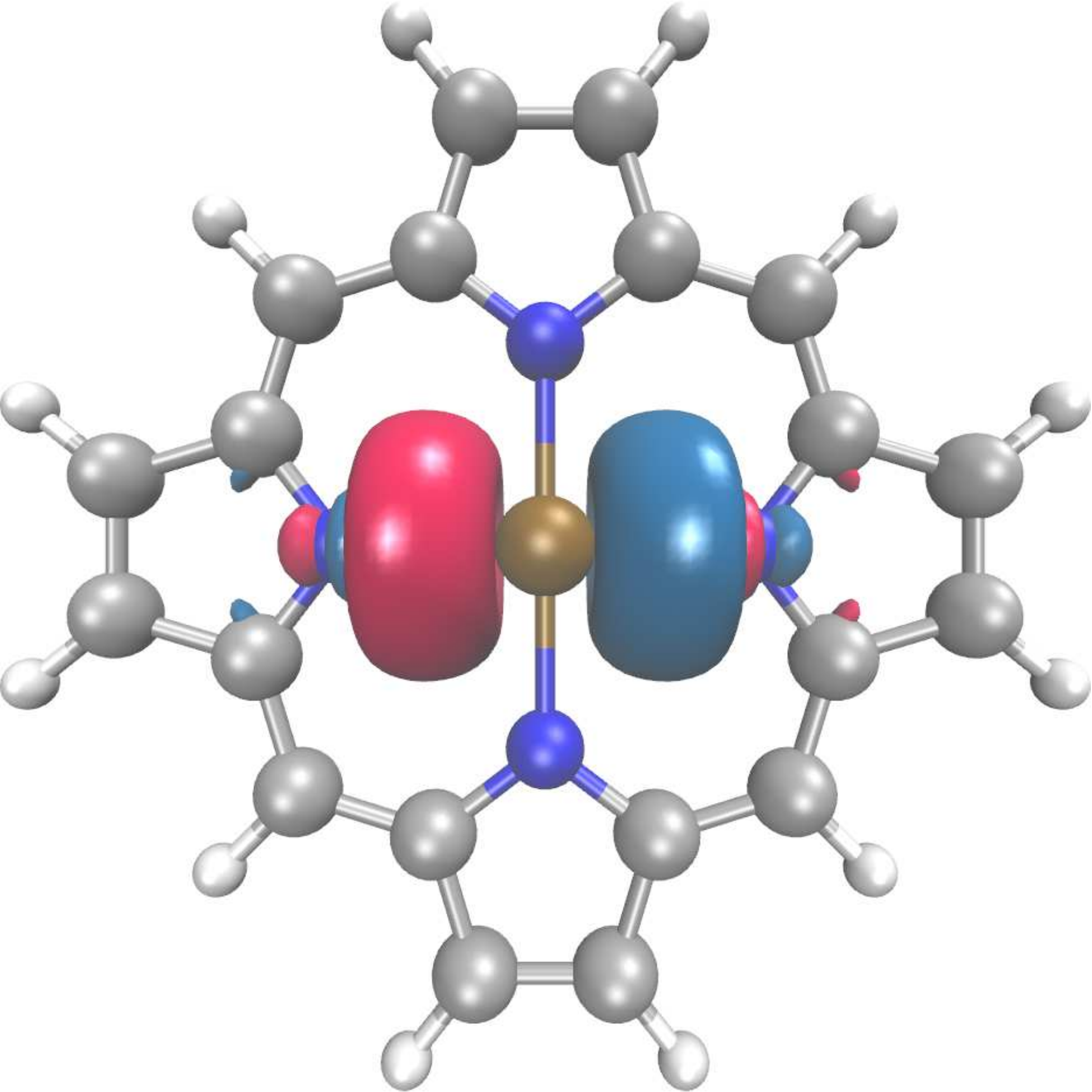}
\subcaption{$e_{u}$}
\end{minipage}
\begin{minipage}{0.24\textwidth}
\includegraphics[width=\linewidth]{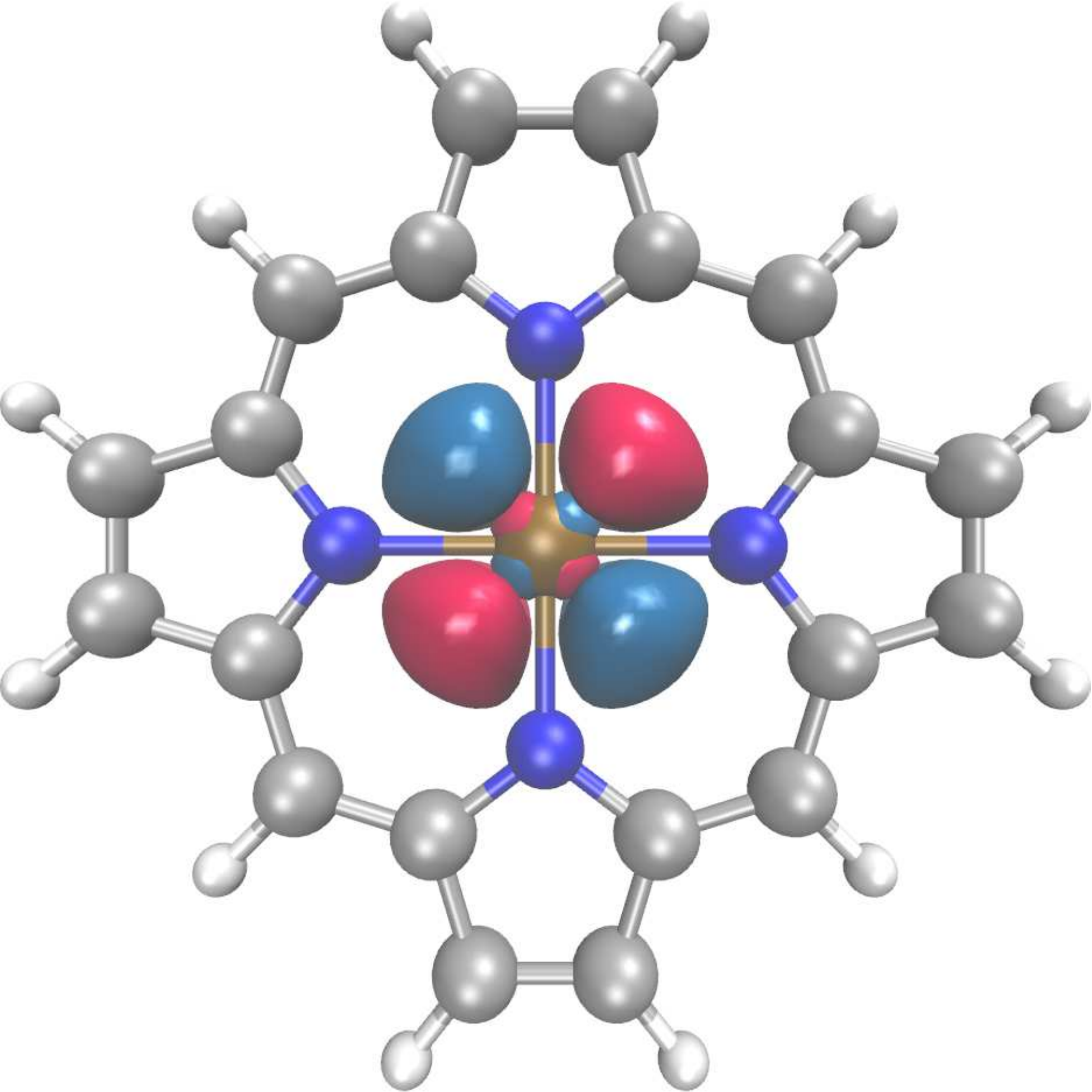}
\subcaption{$b_{2g}$}
\end{minipage}
\caption{Some ASCI-SCF orbitals (without a nodal plane at the molecular plane), for the $^3A_{2g}$ state optimized with the (40e, 42o) active space.}
\label{fig:sigmabonding}
\end{figure}

We consequently examined a more "physical" active space of 40 electrons in 42 orbitals. This consists of the 24 porphyrin $\pi$ orbitals, the $3d$, $4s$, $4p$, and $4d$ shells of Fe, and the four $\sigma$ lone pairs used by N to coordinate to Fe. This active space differs from the (44e, 44o) active space in that the N $2p$ orbitals not used to coordinate to the metal are ignored, while the Fe $4s$ and $4p_z$ are added in to create a complete shell. ASCI-SCF calculations with the (40e, 42o) space preserved orbital symmetry and did not lead to any "slipping" of active orbitals out of the active space, indicating that this was a reasonable and balanced active space. Some representative resulting orbitals have been depicted in Fig \ref{fig:sigmabonding}. The remaining orbitals are available in the checkpoint files in the Supporting Information.

The energies of the stable, low-lying states of Fe(II) porphyrin with this active space are given in Table \ref{tab:actualAS}. A triplet ground state ($^3A_{2g}$) is predicted by this active space, which is consistent with experiment. The PT2 correction proves to be absolutely critical for this ordering, as the $^5A_{1g}$ state appeared to be the lowest in energy at the ASCI-SCF variational level. Interestingly, ASCI-SCF with this smaller active space predicts energies that are below the HCISCF values for the (44e, 44o) active space (both when comparing variational energies and extrapolated ones). While our new active space is not strictly a subspace of the (44e, 44o) space due to the inclusion of the Fe $4s$ and $4p_z$ orbitals, the lower energy values certainly indicate that a more optimal choice of orbitals was reached with this smaller active space, relative to the HCISCF results. Finally, alternate versions of the ASCI algorithm which utilize somewhat less accurate search algorithms (via tuning the search parameter\cite{tubman2016-1,tubman_modern_2018}) appeared to lead to the same solution for the $^3A_{2g}$ state, suggesting that the overall accuracy of the ASCI wave function was not a major issue for this case. This is encouraging with regards to applicability of SCI for this problem, although identical behavior should not be universally expected, and the most accurate wave function (subject to size constraints) should be used whenever possible to minimize error.

\begin{table}[htb!]
\begin{tabular}{lllll}
\hline State      & ASCI-SCF & Extrapolated ASCI+PT2 & E$_\textrm{relative}$(ASCI+PT2) & E$_\textrm{relative}$(PBE0) \\\hline
$^3A_{2g}$ &-2245.1671& -2245.2208(1) & 0.0             & 0.0 \\
$^3E_{g}$  &-2245.1610& -2245.2187(1) & 1.3(1)         & 3.9 \\
            & &                &                 &     \\
$^5A_{1g}$ &-2245.1699 & -2245.2137(2)  & 4.5(1)          & 6.5 \\
$^5E_{g}$  &-2245.1622 & -2245.2062(2)  & 9.2(1)          & 9.7\\\hline
\end{tabular}
\caption{ASCISCF (with 500000 determinants) and extrapolated E$_\text{ASCI}$+E$_\text{PT2}$ (using ASCI-SCF orbitals) for Fe-porphyrin/cc-pVDZ, in a.u.. Relative energies (with respect to the lowest energy $^3A_{2g}$ state) in kcal/mol are also reported for both extrapolated ASCI+PT2 and the PBE0 DFT functional.}
\label{tab:actualAS}
\end{table}

We also examined low lying states of different spatial symmetries with this active space, using the DFT results of \citeref{groenhof2005electronic} as a guide for reasonable $d$ orbital fillings for various spatial symmetries (relative energies from the PBE0 DFT functional are also given in \ref{tab:actualAS}). Restricted open-shell PBE0\cite{pbe0} orbitals were converged with the square gradient minimization (SGM) orbital optimizer\cite{hait2019excited} to create initial guesses with correct spatial symmetries for ASCI-SCF, without risk of collapse to the DFT global minimums of $^5A_{1g}$ and $^3A_{2g}$. However, the current energy gradient based ASCI-SCF orbital optimization could only converge the $^5A_{1g}$, $^5E_{g}$, $^3A_{2g}$ and $^3E_{g}$ states. It therefore appears that these four alone are stable with respect to orbital rotation, as alternative states like $^5B_{2g}$ collapsed into one of them during the course of the optimization. SGM type optimizers for arbitrary extrema in conjunction with ASCI-SCF could therefore prove quite effective in fully mapping out low energy spectra of bioinorganic species like metal porphyrins. For this work, however, we were principally interested in the ground electronic states, and did not examine the behavior of unstable excited states further.

It is also worth noting that the natural orbital occupations (provided in the supporting information) for all four states do not suggest presence of multireference character, and neither does a cluster decomposition\cite{lehotola2017} of the ASCI wave functions. It therefore appears that the model Fe porphyrin is likely not a strongly correlated system, and the larger active space only served to include more dynamical correlation. The effects of out-of-active-space dynamical correlation therefore must be considered for obtaining quantitative energy gaps\cite{li2019role}. Even single reference methods ought to be fairly effective in describing the ground state of Fe porphyrin, although the low spacing between electronic states predicted by many methods\cite{lee2020utilizing,groenhof2005electronic} indicate that the correctly estimating the precise ordering low lying states of various symmetries could well be a challenging task.

\subsection{Strategies for avoiding local extrema}\label{extrema}
Nearly all orbital-optimization based quantum chemistry approaches have the potential to converge to a local extremum instead of the global minimum. MCSCF methods are particularly susceptible to this problem, as there are three sets of available degrees of freedom (orbital rotation, list of determinants treated variationally, and coefficients of said determinants) with substantial levels of linear dependencies between them. For instance, the effect of swapping two active space orbitals could also be realized by adjusting the set of determinants in the variational wave function and their coefficients. ASCI-SCF (or any other selected CI approach) typically exhibits far more local minima than CASSCF, because the energy depends at least weakly on the active-active orbital rotations unlike CASSCF.

These local extrema can pose a significant challenge when ground state solutions are desired, as can be seen from the higher energy HCISCF solutions for porphyrin. Such spurious solutions possess considerable capacity to mislead, especially when dynamical correlation out of the active space is not considered. The compactness of the ASCI wave function assists somewhat in avoiding such extrema, but we have nonetheless encountered a fair few local extrema over the course of our investigations. We have found the following strategies to be useful with regards to obtaining the lowest energy ASCI-SCF solution (although we cannot prove that such solutions are indeed the global minima):

\begin{figure}[htb!]
\includegraphics[width=0.5\textwidth]{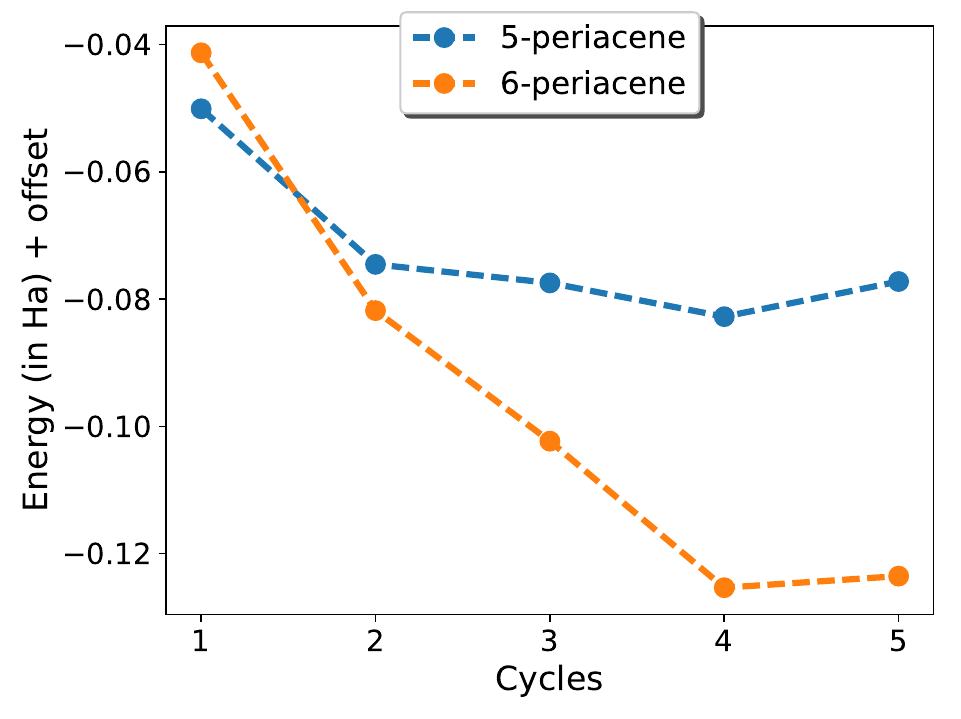}
\caption{Convergence of ASCI-SCF variational energy of singlet periacenes (using 1 million variational determinants for orbital optimization, see point 2 below for details) using the strategy described below to avoid local minima. The absolute energies have been shifted by offsets to fit both species in the same plot. In this case, cycle 4 is the lowest energy solution for both species.}
\label{fig:conv}
\end{figure}

\begin{enumerate}
    \item Use a fairly compact (as accurate as possible for the given size) initial SCI wave function to avoid inclusion of relatively unimportant determinants (and exclusion of more important configurations) that can mislead the orbital optimizer.  The use of natural orbital rotations and more effective selection rules like ASCI can be of great help here.
    \item Converge orbitals first with a small number of determinants, and then feed those orbitals into a larger calculation. Do not recycle the determinant list from the small calculation, but force the SCI solver to recalculate the list of important determinants to prevent the `memory' of the initial calculation from biasing the process. We have found a sequence of $10^5,2.5\times 10^5$ and $10^6$ determinants to be useful in this regard. However, natural orbital rotations are probably best avoided from this point onwards to preserve useful active-active rotations that compactify the wave function. \label{step:converge}
    \item Using these orbitals as an initial guess, calculate orbitals for an adjacent spin-state (triplet for a target singlet, singlet or quintet for a desired triplet etc.). A full-size calculation is probably not necessary, and about $10^5$ determinants will likely suffice.\label{step:adjacent} 
    \item Use these orbitals as the starting guess for a new calculation for the desired spin state (as described in step \ref{step:converge}). \label{step:targeted}.
    \item Repeat steps \ref{step:adjacent}-\ref{step:targeted} until the variational energy ceases to substantially decrease. 
    \item Choose orbitals corresponding to the minimum energy solution out of all the calculations. 
\end{enumerate}

Examples of the efficacy of the aforementioned strategy for the singlet states of 5- and 6-periacenes are demonstrated in Fig \ref{fig:conv}. Use of repeated cycles (i.e. switching between singlet and triplet solution, steps 3 and 4 above) of ASCI-SCF led to a substantial lowering of the variational energy ($0.03$ a.u. for 5-periacene and $0.08$ a.u. for 6-periacene) by the fourth cycle of runs. The fifth cycle however led to an increase in variational energy for both species, and further iterations were subsequently not considered.  

\section{Conclusions}
In this paper, we have described a method to treat medium-sized molecules with active spaces several times larger than is possible with traditional methods. By taking advantage of an extremely efficient ASCI FCI solver, we are able to generate CASSCF quality results for species relevant to biological, inorganic, and organic systems. We have also showed that the use of active-active orbital rotations can substantially improve the compactness of the ASCI wave function, possibly improving the variational energy by more than a tripling of the wave function size. The ability to routinely study large, correlated systems will permit new lines of investigation into the hardest class of electronic structure problems. It would also be desirable to develop ways to cheaply estimate out-of-active-space dynamical correlation for large ASCI-SCF calculations to obtain quantitative predictions that can be directly compared to experiment. In addition, ASCI-SCF can be combined with the SGM orbital optimizer\cite{hait2019excited} to specifically target excited states at CASSCF level of theory. Work along these directions is presently in progress.

\section*{Acknowledgements}
This material is based upon work supported by the U.S. Department of Energy, Office of Science, Office of Advanced Scientific Computing Research and Office of Basic Energy Sciences, Scientific Discovery through Advanced Computing (SciDAC) program. Computational resources provided by the Extreme Science and Engineering Discovery Environment (XSEDE), which is supported by the National Science Foundation Grant No. OCI-1053575, are gratefully acknowledged. N.M.T. is grateful for support from NASA Ames Research Center and support from the AFRL Information Directorate under Grant No. F4HBKC4162G001. S.L. has been supported by Suomen Akatemia (Academy of Finland) through project number 311149. The views and conclusions contained herein are those of the authors and should not be interpreted as necessarily representing the official policies or endorsements, either expressed or implied of AFRL or the U.S. Government. The U.S. Government is authorized to reproduce and distribute reprints for Governmental purpose notwithstanding any copyright annotation thereon.

\section*{Supporting Information}
Further data tables, a discussion of convergence of energy and orbitals with number of determinants, checkpoint files containing orbitals, and sample input files are available in the Supporting Information.

%\newpage
\bibliography{refs}
\end{document}

\begin{center}\LARGE{Supporting Information}\end{center}
\begin{section}{G1 Test Set}
G1 test set in double and triple zeta basis sets with all electrons and orbitals included in the active space, that is, a full CI calculation.
\begin{table}
\scriptsize
\begin{tabular}{cccccc}
&\multirow{2}{*}{ASCI (100k)}&\multirow{2}{*}{ASCI-SCF (100k)}&\multirow{2}{*}{ASCI (300k)}&SCF stab.&Incr. det. stab.\\
&&&&(kcal/mol)&(kcal/mol)\\
H3COH&-115.40054&-115.41209&-115.40857&7.24&5.03\\
CH3Cl&-499.42603&-499.43680&-499.43185&6.75&3.65\\
H3CSH&-438.03796&-438.04838&-438.04498&6.54&4.41\\
HOCl&-535.22099&-535.23033&-535.22647&5.86&3.43\\
SO2&-547.68359&-547.69205&-547.69587&5.31&7.70\\
N2H4&-111.54561&-111.55333&-111.55379&4.85&5.13\\
H2O2&-151.17999&-151.18557&-151.18564&3.50&3.55\\
ClO&-534.57320&-534.57797&-534.57778&2.99&2.87\\
ClF&-559.19411&-559.19829&-559.19771&2.62&2.26\\
H2CO&-114.21265&-114.21653&-114.21694&2.43&2.69\\
C2H6&-79.55671&-79.55990&-79.56623&2.00&5.97\\
SO&-472.66310&-472.66616&-472.66763&1.92&2.85\\
S2&-795.33304&-795.33590&-795.33839&1.79&3.35\\
Si2H6&-581.59617&-581.59893&-581.60259&1.73&4.03\\
CS&-435.60590&-435.60861&-435.60934&1.70&2.16\\
P2&-681.73137&-681.73384&-681.73467&1.55&2.07\\
NO&-129.59714&-129.59852&-129.59949&0.87&1.47\\
HCO&-113.57058&-113.57185&-113.57485&0.80&2.68\\
O2&-149.98481&-149.98607&-149.98652&0.79&1.08\\
SiO&-364.08585&-364.08706&-364.08810&0.76&1.42\\
HCN&-93.18928&-93.19024&-93.19163&0.60&1.47\\
CO2&-188.12887&-188.12977&-188.13731&0.56&5.30\\
Si2&-577.93675&-577.93762&-577.93802&0.55&0.80\\
CO&-113.05586&-113.05652&-113.05695&0.41&0.68\\
Cl2&-919.26435&-919.26495&-919.26932&0.37&3.12\\
PH3&-342.64268&-342.64305&-342.64375&0.23&0.67\\
C2H4&-78.34743&-78.34777&-78.35217&0.21&2.97\\
NH3&-56.40334&-56.40368&-56.40392&0.21&0.36\\
CN&-92.48687&-92.48717&-92.48788&0.19&0.64\\
SiH4&-291.39761&-291.39790&-291.39869&0.18&0.68\\
SiH3&-290.75367&-290.75394&-290.75423&0.17&0.35\\
NaCl&-621.59504&-621.59527&-621.59528&0.15&0.15\\
N2&-109.27820&-109.27842&-109.27892&0.14&0.45\\
F2&-199.09764&-199.09785&-199.09832&0.13&0.43\\
H2S&-398.87100&-398.87118&-398.87132&0.11&0.20\\
PH2&-342.01506&-342.01520&-342.01522&0.09&0.10\\
SiH2 (triplet)&-290.10095&-290.10108&-290.10103&0.09&0.05\\
C2H2&-77.11099&-77.11111&-77.11274&0.08&1.10\\
SiH2 (singlet)&-290.14370&-290.14381&-290.14379&0.07&0.06\\
H2O&-76.24323&-76.24332&-76.24329&0.06&0.04\\
LiF&-107.15761&-107.15770&-107.15763&0.06&0.02\\
NH2&-55.73491&-55.73498&-55.73493&0.04&0.01\\
HF&-100.23014&-100.23020&-100.23012&0.04&-0.01\\
CH4&-40.38843&-40.38849&-40.38903&0.04&0.38\\
HCl&-460.26029&-460.26033&-460.26035&0.03&0.04\\
CH3&-39.71813&-39.71816&-39.71822&0.02&0.06\\
CH2 (singlet)&-39.02464&-39.02468&-39.02465&0.02&0.00\\
OH&-75.56149&-75.56152&-75.56149&0.02&0.00\\
CH2 (triplet)&-39.04353&-39.04355&-39.04353&0.02&0.00\\
NH&-55.09347&-55.09349&-55.09348&0.01&0.01\\
Na2&-323.73402&-323.73404&-323.73403&0.01&0.00\\
CH&-38.38179&-38.38180&-38.38179&0.00&0.00\\
Li2&-14.90133&-14.90134&-14.90134&0.00&0.00\\
LiH&-8.01471&-8.01471&-8.01471&0.00&0.00\\
BeH&-15.18927&-15.18927&-15.18927&0.00&0.00
\end{tabular}
\normalsize
\caption{ASCI and ASCI-SCF variational energies with 100000 determinants and ASCI variational energies with 300000 determinants for the G1 test set (in Ha) in the cc-pvdz basis, sorted by the magnitude of the SCF stabilization. The SCF stabilization and stabilization obtained by tripling the number of determinants (in kcal/mol) is given in the last two column.}
\label{table:DZ}
\end{table}

\begin{table}
\scriptsize
\begin{tabular}{cccccc}
&\multirow{2}{*}{ASCI (100k)}&\multirow{2}{*}{ASCI-SCF (100k)}&\multirow{2}{*}{ASCI (300k)}&SCF stab.&Incr. det. stab.\\
&&&&(kcal/mol)&(kcal/mol)\\
H3CSH&-438.13456&-438.18038&-438.16427&28.75&18.65\\
CO2&-188.26289&-188.30784&-188.29965&28.20&23.06\\
SO2&-547.88864&-547.92974&-547.92515&25.79&22.91\\
CH3Cl&-499.54311&-499.58109&-499.56638&23.83&14.60\\
N2H4&-111.64221&-111.67997&-111.67282&23.69&19.21\\
Si2H6&-581.71766&-581.75427&-581.74293&22.97&15.86\\
C2H6&-79.62516&-79.66140&-79.65307&22.74&17.51\\
H3COH&-115.51465&-115.54928&-115.53570&21.73&13.21\\
Cl2&-919.42110&-919.44838&-919.44523&17.12&15.14\\
HOCl&-535.38388&-535.41065&-535.40107&16.80&10.79\\
ClO&-534.73118&-534.75322&-534.75092&13.83&12.39\\
P2&-681.83866&-681.85976&-681.85962&13.24&13.16\\
ClF&-559.38152&-559.40175&-559.39619&12.70&9.21\\
H2O2&-151.32851&-151.34654&-151.34623&11.31&11.12\\
H2CO&-114.32288&-114.34089&-114.33844&11.30&9.76\\
HCO&-113.67500&-113.69055&-113.68797&9.75&8.14\\
S2&-795.47148&-795.48658&-795.48953&9.48&11.33\\
SO&-472.81489&-472.82774&-472.82764&8.07&8.00\\
O2&-150.12432&-150.13590&-150.13615&7.27&7.43\\
C2H4&-78.43253&-78.44396&-78.44542&7.17&8.09\\
SiO&-364.23609&-364.24745&-364.24694&7.13&6.81\\
CS&-435.70656&-435.71590&-435.71839&5.86&7.43\\
F2&-199.29225&-199.30154&-199.30202&5.83&6.13\\
NO&-129.71736&-129.72627&-129.72686&5.59&5.96\\
HCN&-93.28022&-93.28869&-93.28873&5.32&5.34\\
Si2&-578.06363&-578.07193&-578.07253&5.21&5.58\\
SiH4&-291.48123&-291.48847&-291.48680&4.55&3.50\\
CN&-92.57368&-92.57961&-92.58052&3.72&4.30\\
N2&-109.38184&-109.38740&-109.38802&3.49&3.87\\
SiH3&-290.83343&-290.83879&-290.83792&3.36&2.82\\
C2H2&-77.19895&-77.20407&-77.20463&3.21&3.56\\
CO&-113.16448&-113.16894&-113.16958&2.80&3.20\\
NH3&-56.47972&-56.48362&-56.48319&2.45&2.18\\
NaCl&-621.71441&-621.71824&-621.71727&2.41&1.80\\
LiF&-107.28828&-107.29157&-107.28930&2.06&0.64\\
PH3&-342.72404&-342.72714&-342.72878&1.94&2.97\\
HCl&-460.36396&-460.36582&-460.36637&1.17&1.51\\
SiH2 (singlet)&-290.22385&-290.22564&-290.22533&1.13&0.93\\
SiH2 (triplet)&-290.17857&-290.18021&-290.18005&1.03&0.93\\
H2O&-76.34121&-76.34280&-76.34306&1.00&1.16\\
H2S&-398.96280&-398.96426&-398.96584&0.91&1.90\\
PH2&-342.09331&-342.09464&-342.09550&0.84&1.38\\
HF&-100.34859&-100.34968&-100.34981&0.68&0.76\\
CH2 (singlet)&-39.07391&-39.07448&-39.07463&0.36&0.45\\
NH2&-55.80487&-55.80540&-55.80603&0.33&0.72\\
CH2 (triplet)&-39.07362&-39.07398&-39.09185&0.22&11.43\\
NH&-55.15224&-55.15258&-55.15259&0.21&0.22\\
OH&-75.64903&-75.64936&-75.64952&0.20&0.31\\
CH4&-40.44945&-40.44974&-40.45163&0.18&1.37\\
Na2&-323.76884&-323.76911&-323.76901&0.17&0.10\\
CH3&-39.77454&-39.77467&-39.77556&0.08&0.64\\
CH&-38.42199&-38.42205&-38.42208&0.04&0.06\\
BeH&-15.20306&-15.20306&-15.20306&0.00&0.00\\
Li2&-14.93078&-14.93078&-14.93078&0.00&0.00\\
LiH&-8.03648&-8.03648&-8.03648&0.00&0.00
\end{tabular}
\normalsize
\caption{ASCI and ASCI-SCF variational energies with 100000 determinants and ASCI variational energies with 300000 determinants for the G1 test set (in Ha) in the cc-pvtz basis, sorted by the magnitude of the SCF stabilization. The SCF stabilization and stabilization obtained by tripling the number of determinants (in kcal/mol) is given in the last two column.}
\label{table:TZ}
\end{table}
\end{section}
\newpage
\section{DFT results for periacenes}
% \usepackage{booktabs}
\begin{table}[]
\begin{tabular}{@{}lllllll@{}}
\toprule
Species     & v2RDM & ASCI    & SPW92 & BLYP & B97M-V & B3LYP \\ \midrule
2-periacene & 35.4  & 38.3(1) & 35.2  & 32.6 & 33.2   & 34.2  \\
3-periacene           & 19.5  & 17.5(2) & 14.5  & 12.8 & 11.3   & 12.8  \\
4-periacene           & 13.4  & 10.0(6) & 3.8   & 3.8  & 5.0    & 6.2   \\
5-periacene           & 10.8  & 4 (1)   & 0.9   & 1.1  & 3.1    & 4.7   \\
6-periacene           & 8.9   & 8 (1)   & 0.9   & 1.3  & 4.5    & 7.6   \\ \bottomrule
\end{tabular}
\caption{Comparison of singlet-triplet gaps (in kcal/mol) for various quantum chemistry techniques, using geometries from \citeref{mullinax2019heterogeneous} and the cc-pVDZ basis set. The DFT singlet energies were spin-purified via Yamaguchi's approximate projection method for accuracy in dealing with biradical species.}
\label{tab:my-table}
\end{table}

\section{Convergence of energies and orbitals with number of determinants}

In selected CI calculations, we attempt to obtain Full CI quality results from only a subset of the Full CI wave function. In the case of approximating a CASSCF wave function with selected CI, we must answer two questions: "how converged is the energy to the full CI results?" and "how converged are the orbitals to the CASSCF result?"

\begin{figure}[htb!]
\includegraphics[width=0.5\textwidth]{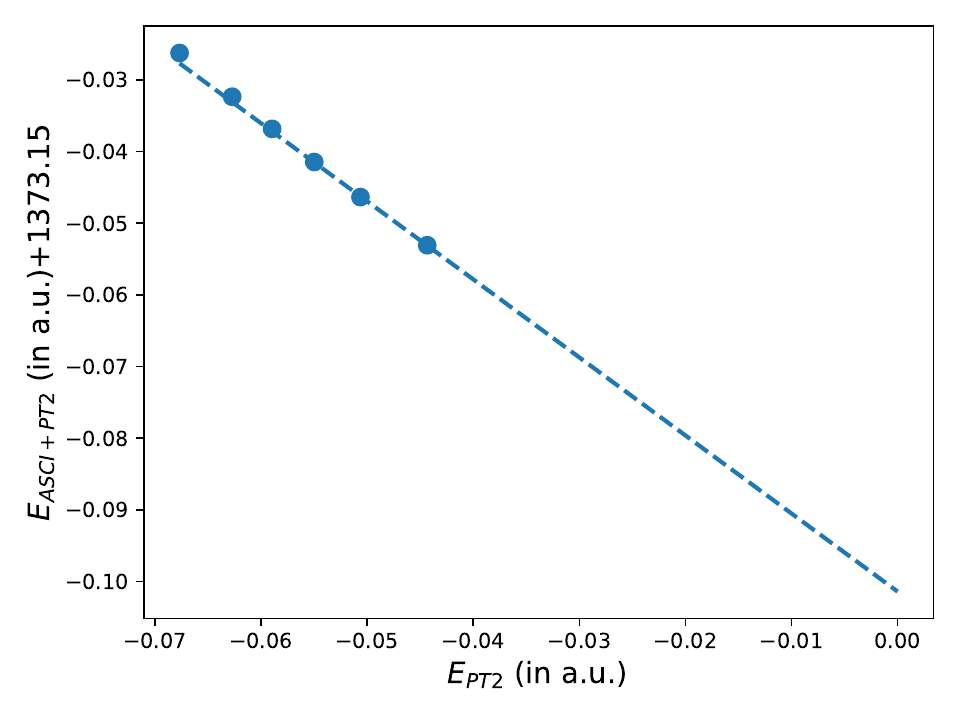}
\caption{Linear extrapolation to the CASCI limit for singlet 4-periacene. Only the last three points (corresponding to 1, 2 and 5 million determinants) were used in the extrapolation, but the other points presented in Tab \ref{tab:si-4-periacene} were added to show general trend. }
\label{fig:conv}
\end{figure}

As noted in the main body, all results presented utilize an extrapolation of the ASCI+PT2 energy to the limit where the PT2 contribution is zero. As the selected CI approaches the Full CI result in the variational space, the contribution of the PT2 correction shrinks to this limit. This extrapolation is critical to obtaining highly accurate energies and relative energies. A plot of the ASCI+PT2 energy vs. the PT2 component is very close to linear, allowing the easy estimation of the true FCI result (as can be seen from the example in Fig \ref{fig:conv}). Typically, we use 1 million, 2 million, and 5 million determinant results to make this extrapolation. Further details of the extrapolation are given in \citeref{hait2019levels}.

\begin{table}[]
\begin{tabular}{@{}lcccc@{}}
\toprule
         & \multicolumn{2}{c}{4-periacene} & \multicolumn{2}{c}{Fe porphyrin}\\
NDETS    & Singlet & Triplet & $^3A_{2g}$ & $^5A_{1g}$ \\ \midrule
100000   & -1373.176251  & -1373.175692 & -2245.189127 & -2245.187915 \\
250000   & -1373.182348  & -1373.183468 & -2245.193157 & -2245.191480 \\
500000   & -1373.186834  & -1373.189043 & -2245.197974 & -2245.195652 \\
1000000  & -1373.191459  & -1373.194255 & -2245.200969 & -2245.198219 \\
2000000  & -1373.196352  & -1373.199033 & -2245.203383 & -2245.200280 \\
5000000  & -1373.203069  & -1373.204548 & -2245.205987 & -2245.202529 \\
extrapolated& -1373.251386  & -1373.235395 & -2245.220769 & -2245.213680 \\ \bottomrule
\end{tabular}
\caption{4-periacene (orbitals determined with 1 million determinants) and Fe porphyrin (orbitals determined with 500000 determinants) ASCI+PT2 energies with wave functions of various sizes.}
\label{tab:si-4-periacene}
\end{table}

That said, one may ask how these ASCI "single point" energies change with various determinant sizes. Table \ref{tab:si-4-periacene} contains the energies of the singlet and triplet states of 4-periacene and Fe porphyrin with increasing numbers of determinants. Note that each spin state of each molecule utilizes one common set of orbitals for all wave function sizes, so the results report directly on the convergence of the ASCI method toward the FCI result. We observe that increasing the wave function by approximately a factor of 2 in the larger wave functions used here results in a lowering of the energy by approximately 5 mHa or about 3 kcal/mol for both spin states of 4-periacene and about 2-3 mHa for the spin states of Fe porphyrin. Given the small spin-state gaps in the case of 4-periacene at "single-points", it may be surprising that the extrapolated values are so well separated in energy. Moreover, the magnitude (and in some cases, the sign) of the singlet-triplet gap is not properly estimated unless the extrapolation is performed. We therefore do not advocate using non-extrapolated data unless the wave function has been observed to be very tightly converged (with energy changes considerably less than 0.1 mHa with increasing numbers of determinants). To summarize, while the variation in ASCI+PT2 energies as an increasing number of determinants is used is not very large, one should nonetheless carry out progressively larger calculations until the extrapolated results is converged to the desired accuracy.

\begin{table}[]
\begin{tabular}{@{}lcccc@{}}
\toprule
NDETS (orbital set)    & 4-periacene & 6-periacene \\ \midrule
250000 (100k)   & -1373.117509 & -1981.551219 \\
250000 (250k)   & -1373.118483 & -1981.575160\\
1000000 (250k)  & -1373.134295 & -1981.602070\\
1000000 (1m) & -1373.136459 & -1981.625352\\ 
2000000 (1m) &              & -1981.636391\\
2000000 (2m) &              & -1981.638488\\
5000000 (2m) &              & -1981.652341\\ 
5000000 (5m) &              & -1981.652910\\ \bottomrule
\end{tabular}
\caption{4-periacene and 6-periacene ASCI variational energies with wave functions of various sizes using orbitals determined using wave functions of various sizes. For example, "100k" orbitals are determined by carrying out ASCI-SCF with a wave function with 100000 determinants.}
\label{tab:si-orbconv}
\end{table}

Since an approximate wave function is used to compute orbital gradients in ASCI-SCF, one may also ask whether this approximation leads to a deterioration of the resulting orbitals. We can measure orbital convergence by computing orbitals with wave functions of different size and comparing the variational energies obtained when those orbitals are used with wave functions of the same size. For example, Table \ref{tab:si-orbconv} gives the variational energies of 4- and 6-periacene using various sets of orbitals. We describe an orbital set by the size of wave function used in the ASCI-SCF calculation which produced those orbitals. Hence a "1m" orbital set was obtained by carrying out ASCI-SCF with a wave function with 1 million determinants. What becomes immediately apparent by comparing the first two rows of 4-periacene and the third and fourth rows, is that increasing the size of wave function used to determine the orbitals contributes between 0.5 and 1.5 kcal/mol to the absolute energy; the orbitals obtained with 1 million determinants are not much better than those determined with 250000 determinants or even 100000. On the other hand, the larger 6-periacene system displays larger variations with orbital sets obtained with these smaller wave function sizes. Increasing the size of the wave function used to determine the orbitals lowers the energy by about 15 kcal/mol when going from 100000 to 250000 and the same amount again going from 250000 to 1 million. However, using 2 million or 5 million determinants is again not observed to lead to substantial improvement. It is therefore safe to say that orbitals converge quite readily even though the wave function used to compute the orbital gradients is approximate. As a further technical note, there are two potential sources of stabilization in ASCI-SCF: active-active rotations which improve the convergence of the ASCI wave function by creating a more compact wave function and out-of-active-space rotations (with the inactive or virtual orbitals) which alter the partitioning of the orbital space. The distinction is meaningful because stabilization due to improving the ASCI wave function's accuracy (active-active rotations) can be further improved by the PT2 calculation and extrapolation while the other cannot be so improved. At this time, we do not have a means of determining how much of the additional stabilization is merely active-active rotations and how much is genuine out-of-active-space rotations. In any case, we have shown that the ASCI energies obtained are not strongly dependent on the size of the wave function used to determine the orbitals above a certain size.
\section{Single reference nature of Fe Porphyrin}
Our computations seem to suggest Fe porphyrin is overall fairly single reference. The strongest evidence in favor of this are the active space natural orbital occupations given in Table \ref{tab:porphnoons}, which show very little deviation from the single determinant values of $2$ (for doubly occupied orbitals), $1$ (for singly occupied orbitals) or $0$ for virtual orbitals. The largest deviations from the single determinant values are of the order of $\approx 0.12$ electrons, which is not particularly large (for context, indisputably single-reference ethene in the same cc-pVDZ basis has 0.1 as the largest deviation). It is also important to note that the active space is mostly made of $\pi$ ligand orbitals. Full $\pi$ space calculations for acenes are also known to predict larger multireference character than full valence calculations which include $\sigma$ orbitals\cite{lee2017coupled,lehtola2018orbital}, and so it is entirely possible that the natural orbital occupations for our $\pi$ dominated active space appears to be somewhat more multireference than the actual FCI results for the porphyrin model system studied.

\begin{table}[htb!]
\begin{tabular}{l|lll|ll}
\hline $n$ & \multicolumn{2}{c}{$^5$A$_{1g}$}   &  & \multicolumn{2}{c}{$^3$A$_{2g}$}   \\\hline
  & $|T_n|^2$    & $|T_n|^2/|C_n|^2$ &  & $|T_n|^2$    & $|T_n|^2/|C_n|$ \\\hline 
1 & 4.95E-03 & 1.00E+00    &  & 6.50E-03 & 1.00E+00    \\
2 & 4.27E-01 & 1.00E+00    &  & 4.46E-01 & 1.00E+00    \\
3 & 8.78E-03 & 8.83E-01    &  & 9.14E-03 & 8.84E-01    \\
4 & 1.36E-03 & 2.25E-02    &  & 1.72E-03 & 2.60E-02    \\
5 & 1.70E-05 & 3.78E-02    &  & 1.80E-05 & 3.68E-02    \\
6 & 9.28E-06 & 5.58E-03    &  & 1.45E-05 & 7.57E-03   
\end{tabular}
\caption{Cluster decomposition of the ASCI wave functions for the $^5$A$_{1g}$ and $^3$A$_{2g}$ states of Fe porphyrin (using 5 million determinants, within the (40e,42o) active space). $T_n$ and $C_n$ are the cluster $t$ amplitudes and CI coefficients of order $n$, respectively. The norms $\abs{T_n}$ and $\abs{C_n}$ are $L_2$ norms. For further details see \citeref{lehotola2017}.}
\label{tab:clusterdec}
\end{table}

Another point in support of porphyrin being single reference are the cluster decompositions of the ASCI wave functions (using 5 million determinants and ASCI-SCF orbitals, following the protocol described in \citeref{lehotola2017}) that are supplied in Table \ref{tab:clusterdec}. These show that the most important cluster amplitudes are doubles and triples. The quadruples are about an order of magnitude smaller than the triples and the higher order terms even smaller, suggesting that coupled cluster models that can account for a reasonable description of triples could be fairly successful in describing Fe porphyrin. This is in contrast to truly multireference transition metal systems\cite{hait2019levels,amaya2015} where CCSDTQ can prove insufficient. 

\begin{table}[hb!]
\small
\begin{tabular}{llll}
\hline
$^3$A$_{2g}$ & $^3$E$_{g}$ & $^5$A$_{1g}$ & $^5$E$_{g}$ \\ \hline
0.01        & 0.01       & 0.01        & 0.01       \\
0.01        & 0.01       & 0.01        & 0.01       \\
0.01        & 0.01       & 0.01        & 0.01       \\
0.01        & 0.01       & 0.01        & 0.01       \\
0.01        & 0.01       & 0.01        & 0.01       \\
0.01        & 0.01       & 0.01        & 0.01       \\
0.01        & 0.01       & 0.01        & 0.01       \\
0.02        & 0.02       & 0.01        & 0.01       \\
0.03        & 0.02       & 0.03        & 0.02       \\
0.03        & 0.03       & 0.03        & 0.03       \\
0.03        & 0.03       & 0.03        & 0.03       \\
0.03        & 0.03       & 0.03        & 0.03       \\
0.03        & 0.03       & 0.03        & 0.03       \\
0.04        & 0.04       & 0.04        & 0.04       \\
0.04        & 0.04       & 0.05        & 0.05       \\
0.04        & 0.04       & 0.05        & 0.05       \\
0.05        & 0.04       & 0.05        & 0.05       \\
0.05        & 0.05       & 0.07        & 0.07       \\
0.07        & 0.07       & 0.12        & 0.13       \\
0.12        & 0.12       & 0.13        & 0.13       \\
0.12        & 0.12       & 1.00        & 0.99       \\
0.99        & 1.05       & 1.00        & 1.00       \\
0.99        & 1.05       & 1.00        & 1.00       \\
1.90        & 1.90       & 1.00        & 1.00       \\
1.91        & 1.91       & 1.89        & 1.89       \\
1.94        & 1.92       & 1.91        & 1.91       \\
1.95        & 1.92       & 1.94        & 1.94       \\
1.95        & 1.94       & 1.94        & 1.94       \\
1.95        & 1.95       & 1.94        & 1.94       \\
1.96        & 1.95       & 1.95        & 1.95       \\
1.96        & 1.95       & 1.95        & 1.95       \\
1.97        & 1.96       & 1.95        & 1.95       \\
1.97        & 1.96       & 1.96        & 1.96       \\
1.97        & 1.97       & 1.97        & 1.97       \\
1.97        & 1.97       & 1.98        & 1.97       \\
1.98        & 1.97       & 1.98        & 1.98       \\
1.98        & 1.98       & 1.98        & 1.98       \\
1.98        & 1.98       & 1.98        & 1.98       \\
1.98        & 1.98       & 1.98        & 1.98       \\
1.98        & 1.98       & 1.98        & 1.98       \\
1.98        & 1.98       & 1.98        & 1.98       \\
1.99        & 1.99       & 1.99        & 1.99       \\ \hline
\end{tabular}
\normalsize
\caption{Natural orbital occupations of electronic states of porphyrin (as estimated from ASCI wave functions with 5 million determinants, within the (40e,42o) active space).}
\label{tab:porphnoons}
\end{table}

\newpage
\bibliography{refs}